\documentclass[ twocolumn]{aastex62}
\graphicspath{{./}{figures/}}

\shorttitle{\nth{1} \emph{Insight-}HXMT Gamma-Ray Burst Catalog}
\shortauthors{X. Y. Song et al.}

\usepackage{hyperref}
\usepackage{csquotes}
\usepackage{graphicx}
\usepackage{epstopdf}
\usepackage{lineno}
\usepackage{nicefrac}
\usepackage{amsmath,amstext,mathrsfs}
\usepackage[super]{nth}
\usepackage{comment}
\usepackage{xspace}
\usepackage{soul}
\usepackage{xfrac}

\begin{document}

\title{The First \emph{Insight-}HXMT Gamma-Ray Burst Catalog: The First Four Years}
\correspondingauthor{Shao-Lin Xiong, Shuang-Nan Zhang}
\email{xiongsl@ihep.ac.cn, zhangsn@ihep.ac.cn}
\author{Xin-Ying Song}
\affil{Key Laboratory of Particle Astrophysics, Institute of High Energy Physics, Chinese Academy of Sciences, Beijing 100049, China}
\author{Shao-Lin Xiong}
\affil{Key Laboratory of Particle Astrophysics, Institute of High Energy Physics, Chinese Academy of Sciences, Beijing 100049, China}
\author{Shuang-Nan Zhang}
\affil{Key Laboratory of Particle Astrophysics, Institute of High Energy Physics, Chinese Academy of Sciences, Beijing 100049, China}
\affil{University of Chinese Academy of Sciences, Chinese Academy of Sciences, Beijing 100049, China}
\author{Cheng-Kui Li}
\affil{Key Laboratory of Particle Astrophysics, Institute of High Energy Physics, Chinese Academy of Sciences, Beijing 100049, China}
\author{Xiao-Bo Li}
\affil{Key Laboratory of Particle Astrophysics, Institute of High Energy Physics, Chinese Academy of Sciences, Beijing 100049, China}
\author{Yue Huang}
\affil{Key Laboratory of Particle Astrophysics, Institute of High Energy Physics, Chinese Academy of Sciences, Beijing 100049, China}

\author{Cristiano Guidorzi}
\affil{Department of Physics and Earth Science, University of Ferrara, Via
Saragat 1, I-44122 Ferrara, Italy}
\affil{INFN - Sezione di Ferrara, Via Saragat 1, I-44122 Ferrara, Italy}
\affil{INAF - Osservatorio di Astrofisica e Scienza dello Spazio di
Bologna, Via Piero Gobetti 101, I-40129 Bologna, Italy}

\author{Filippo Frontera}
\affil{Department of Physics and Earth Science, University of Ferrara, Via
Saragat 1, I-44122 Ferrara, Italy}
\affil{INAF - Osservatorio di Astrofisica e Scienza dello Spazio di
Bologna, Via Piero Gobetti 101, I-40129 Bologna, Italy}

\author{Cong-Zhan Liu}
\affil{Key Laboratory of Particle Astrophysics, Institute of High Energy Physics, Chinese Academy of Sciences, Beijing 100049, China}
\author{Xu-Fang Li}
\affil{Key Laboratory of Particle Astrophysics, Institute of High Energy Physics, Chinese Academy of Sciences, Beijing 100049, China}
\author{Gang Li}
\affil{Key Laboratory of Particle Astrophysics, Institute of High Energy Physics, Chinese Academy of Sciences, Beijing 100049, China}
\author{Jin-Yuan Liao}
\affil{Key Laboratory of Particle Astrophysics, Institute of High Energy Physics, Chinese Academy of Sciences, Beijing 100049, China}
\author{Ce Cai}
\affil{Key Laboratory of Particle Astrophysics, Institute of High Energy Physics, Chinese Academy of Sciences, Beijing 100049, China}
\affil{University of Chinese Academy of Sciences, Chinese Academy of Sciences, Beijing 100049, China}
\author{Qi Luo}
\affil{Key Laboratory of Particle Astrophysics, Institute of High Energy Physics, Chinese Academy of Sciences, Beijing 100049, China}
\affil{University of Chinese Academy of Sciences, Chinese Academy of Sciences, Beijing 100049, China}
\author{Shuo Xiao}
\affil{Key Laboratory of Particle Astrophysics, Institute of High Energy Physics, Chinese Academy of Sciences, Beijing 100049, China}
\affil{University of Chinese Academy of Sciences, Chinese Academy of Sciences, Beijing 100049, China}
\author{Qi-Bin Yi}
\affil{Key Laboratory of Particle Astrophysics, Institute of High Energy Physics, Chinese Academy of Sciences, Beijing 100049, China}
\affil{Department of Physics, Xiangtan University, Xiangtan, Hunan Province 411105, China}
\author{Yao-Guang Zheng}
\affil{Key Laboratory of Particle Astrophysics, Institute of High Energy Physics, Chinese Academy of Sciences, Beijing 100049, China}
\affil{College of physics Sciences \& Technology, Hebei University, No. 180 Wusi Dong Road, Lian Chi District, Baoding City, Hebei Province 071002, China}
\author{Deng-Ke Zhou}
\affil{Key Laboratory of Particle Astrophysics, Institute of High Energy Physics, Chinese Academy of Sciences, Beijing 100049, China}
\affil{University of Chinese Academy of Sciences, Chinese Academy of Sciences, Beijing 100049, China}
\author{Jia-Cong Liu}
\affil{Key Laboratory of Particle Astrophysics, Institute of High Energy Physics, Chinese Academy of Sciences, Beijing 100049, China}
\affil{University of Chinese Academy of Sciences, Chinese Academy of Sciences, Beijing 100049, China}
\author{Wang-Chen Xue}
\affil{Key Laboratory of Particle Astrophysics, Institute of High Energy Physics, Chinese Academy of Sciences, Beijing 100049, China}
\affil{University of Chinese Academy of Sciences, Chinese Academy of Sciences, Beijing 100049, China}
\author{Yan-Qiu Zhang}
\affil{Key Laboratory of Particle Astrophysics, Institute of High Energy Physics, Chinese Academy of Sciences, Beijing 100049, China}
\affil{University of Chinese Academy of Sciences, Chinese Academy of Sciences, Beijing 100049, China}
\author{Chao Zheng}
\affil{Key Laboratory of Particle Astrophysics, Institute of High Energy Physics, Chinese Academy of Sciences, Beijing 100049, China}
\affil{University of Chinese Academy of Sciences, Chinese Academy of Sciences, Beijing 100049, China}
\author{Zhi Chang}
\affil{Key Laboratory of Particle Astrophysics, Institute of High Energy Physics, Chinese Academy of Sciences, Beijing 100049, China}
\author{Zheng-Wei Li}
\affil{Key Laboratory of Particle Astrophysics, Institute of High Energy Physics, Chinese Academy of Sciences, Beijing 100049, China}
\author{Xue-Feng Lu}
\affil{Key Laboratory of Particle Astrophysics, Institute of High Energy Physics, Chinese Academy of Sciences, Beijing 100049, China}
\author{Ai-Mei Zhang}
\affil{Key Laboratory of Particle Astrophysics, Institute of High Energy Physics, Chinese Academy of Sciences, Beijing 100049, China}
\author{Yi-Fei Zhang}
\affil{Key Laboratory of Particle Astrophysics, Institute of High Energy Physics, Chinese Academy of Sciences, Beijing 100049, China}
\author{Yong-Jie Jin}
\affil{Department of Engineering Physics, Tsinghua University, Beijing 100084, China}
\author{Ti-Pei Li}
\affil{Key Laboratory of Particle Astrophysics, Institute of High Energy Physics, Chinese Academy of Sciences, Beijing 100049, China}
\affil{Department of Engineering Physics, Tsinghua University, Beijing 100084, China}
\author{Fang-Jun Lu}
\affil{Key Laboratory of Particle Astrophysics, Institute of High Energy Physics, Chinese Academy of Sciences, Beijing 100049, China}
\affil{University of Chinese Academy of Sciences, Chinese Academy of Sciences, Beijing 100049, China}
\author{Li-Ming Song}
\affil{Key Laboratory of Particle Astrophysics, Institute of High Energy Physics, Chinese Academy of Sciences, Beijing 100049, China}
\affil{University of Chinese Academy of Sciences, Chinese Academy of Sciences, Beijing 100049, China}
\author{Mei Wu}
\affil{Key Laboratory of Particle Astrophysics, Institute of High Energy Physics, Chinese Academy of Sciences, Beijing 100049, China}
\author{Yu-Peng Xu}
\affil{Key Laboratory of Particle Astrophysics, Institute of High Energy Physics, Chinese Academy of Sciences, Beijing 100049, China}
\author{Xiang Ma}
\affil{Key Laboratory of Particle Astrophysics, Institute of High Energy Physics, Chinese Academy of Sciences, Beijing 100049, China}

\author{Ming-Yu Ge}
\affil{Key Laboratory of Particle Astrophysics, Institute of High Energy Physics, Chinese Academy of Sciences, Beijing 100049, China}

\author{Shu-Mei Jia}
\affil{Key Laboratory of Particle Astrophysics, Institute of High Energy Physics, Chinese Academy of Sciences, Beijing 100049, China}

\author{Bing Li}
\affil{Key Laboratory of Particle Astrophysics, Institute of High Energy Physics, Chinese Academy of Sciences, Beijing 100049, China}

\author{Jian-Yin Nie}
\affil{Key Laboratory of Particle Astrophysics, Institute of High Energy Physics, Chinese Academy of Sciences, Beijing 100049, China}

\author{Ling-Jun Wang}
\affil{Key Laboratory of Particle Astrophysics, Institute of High Energy Physics, Chinese Academy of Sciences, Beijing 100049, China}

\author{Juan Zhang}
\affil{Key Laboratory of Particle Astrophysics, Institute of High Energy Physics, Chinese Academy of Sciences, Beijing 100049, China}

\author{Shi-Jie Zheng}
\affil{Key Laboratory of Particle Astrophysics, Institute of High Energy Physics, Chinese Academy of Sciences, Beijing 100049, China}

\author{Xue-Juan Yang}
\affil{Department of Physics, Xiangtan University, Xiangtan, Hunan Province 411105, China}

\author{Rong-Jia Yang}
\affil{College of physics Sciences \& Technology, Hebei University, No. 180 Wusi Dong Road, Lian Chi District, Baoding City, Hebei Province 071002, China}

\begin{abstract}
The Hard X-ray Modulation Telescope (\emph{Insight-}HXMT), is China's first X-ray astronomy satellite launched on June 15, 2017. The anti-coincidence CsI detectors of the High Energy X-ray telescope (HE) onboard \emph{Insight-}HXMT could serve as an all-sky gamma-ray monitor in about 0.2-3~MeV. In its first four years of operation, \emph{Insight-}HXMT has detected 322 Gamma-Ray Bursts (GRBs) by offline search pipeline including blind search and targeted search. For the GOLDEN sample of \emph{Insight-}HXMT GRBs, joint analyses were performed with other GRB missions, including \emph{Fermi} Gamma-ray Burst Monitor (\emph{Fermi}/GBM), Swift Burst Alert Telescope (\emph{Swift}/BAT) and Gravitational wave high-energy Electromagnetic Counterpart All-sky Monitor (GECAM). It shows that \emph{Insight-}HXMT can provide better constraint on GRB spectrum at higher energy band. The properties of \emph{Insight-}HXMT GRBs are reported in detail, including their trigger time, duration, spectral parameters, peak fluxes of different time scales and fluence. This catalog is an official product of the \emph{Insight-}HXMT GRB team.
\end{abstract}


\section{Introduction} \label{sec:intro}

The Hard X-ray Modulation Telescope (HXMT), dubbed
\emph{Insight-}HXMT, launched on June 15, 2017, is
China's first X-ray astronomy satellite devoted to broad
band observations in the 1-250~keV band. \emph{Insight-}HXMT consists of three collimator-based telescopes: the High Energy X-ray Telescope \citep[HE,][]{HEofHXMT},
the Medium Energy X-ray Telescope \citep[ME,][]{MEofHXMT} and the Low Energy X-ray Telescope \citep[LE,][]{LEofHXMT}.
\emph{Insight-}HXMT/HE adopts an array of 18 NaI(Tl)/CsI(Na) phoswich
detectors as the main detector plane, with a total geometric
area of about 5100 cm$^{2}$ and a combined Field of
View (FOV) of about $5.7^{\circ} \times 5.7^{\circ}$ (FWHM)\cite[HE,][]{HEofHXMT}. The CsI(Na) (CsI for brevity) in
the phoswich detector of HE can be also used as an all-sky gamma-ray monitor in 0.2-3~MeV, since gamma-rays in this energy range can penetrate the spacecraft and deposit energy in the CsI detecor. Thus HE could play an important role in monitoring gamma-ray bursts (GRBs), MeV pulsars, Solar Flares (SFL), Terrestrial Gamma-ray Flashes (TGF) and
other gamma-ray sources. The measured energy range (deposited energy) of CsI is about 40-800 keV in the normal gain (NG) mode and about 200 keV- 3 MeV in low gain (LG) mode. The NG and LG modes are achieved by adjusting the high voltage of the PMTs which readout the NaI(Tl)/CsI(Na) phoswich detectors.

There is neither on-board GRB trigger system nor prompt data telemetry for \emph{Insight-}HXMT, thus GRBs are unveiled on-ground using two pipelines: the blind search is performed on all data acquired by \emph{Insight-}HXMT/HE CsI detector; in addition, for each GRB detected by other astronomical satellites or instruments (called external triggers hereafter, in contrast to the internal triggers produced by \emph{Insight-}HXMT itself) reported through the Gamma-ray Coordinates Network (GCN), the targeted search pipeline would be launched to search CsI data around the trigger time of external triggers.

From June 26th in 2017 to June 30th 2021, GRBs and others bursts including terrestrial gamma-ray flash (TGF), solar flares \citep[SFL,][]{2021ApJ...918...42Z} have been detected by \emph{Insight-}HXMT, the gamma-ray counterparts of Gravitational Wave (GW) events \citep{2018Insight,pro,2017GCN.21593....1L}; Fast radio bursts \citep[FRB,][]{2020A&A...642A.160G} ) and High Energy Neutrino \citep[HEN, e.g.][]{2020IceCube}, have been monitored as well. During the first four years of observation data, 660 bursts are found, where 322 are classified as GRBs, 245 as TGFs, 4 as SFLs, 33 as charged particles and 56 as unclassified. Targeted search were implemented for external triggers of great importance, including 41 HENs, 48 GWs and 39 FRBs.
For each GRB, the trigger time in Universal Time Coordinated (UTC) and Mission Elapsed Time (MET) are recorded and the event data is selected between $\rm T_{0}-100$s and $\rm T_{0}+300$s in most cases, where $\rm T_{0}$ is trigger time from GCN or blind search. In addition, quick-analysis tools for GRBs and joint spectral fitting with other missions are developed for this catalog analysis, which are introduced in the following sections.

We start with a brief description of the HXMT/CsI detectors and calibration in Section~\ref{sec:instru} and refer~\cite{LUO20201} for a more thorough and complete description of
the instrument and calibration respectively. This is followed, in Section~\ref{sec:algo} the burst-searching algorithm and its up-to-date ability is introduced in details. In Section~\ref{sec:catalogana}, a description of the methodology used in the production of this catalog is presented, including detector selection, data types used, energy selection and background fitting, and the source selection. In Section~\ref{sec:catalog}, catalog analysis and results are presented. In Section~\ref{sec:discussion}, the distribution of duration, hardness are discussed and all the GRBs catalog results are summarized.

\section{Instrument} \label{sec:instru}

\begin{figure*}[!htbp]
  \centering
  \includegraphics[width=0.6\textwidth]{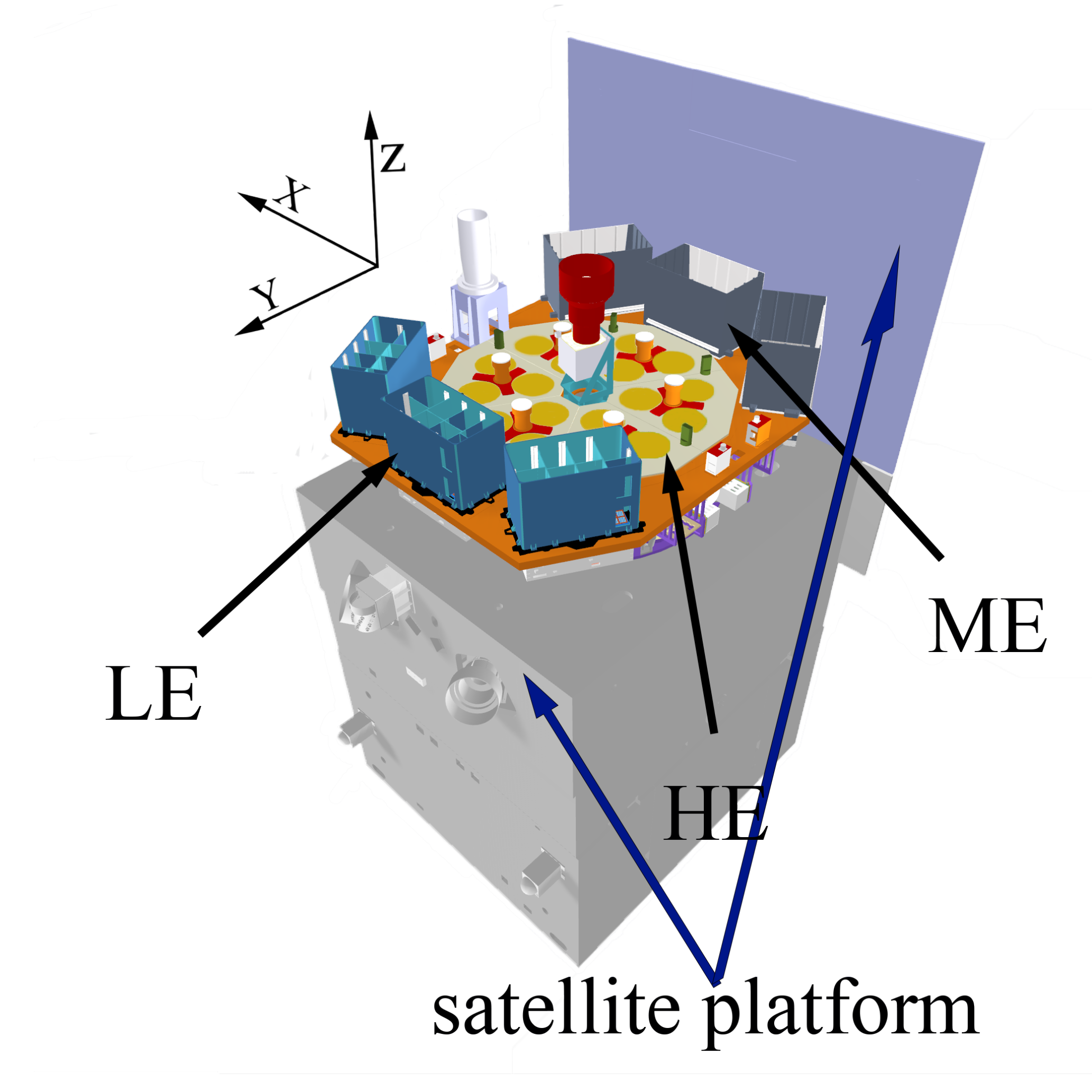}\\
  \caption{The illustration of the satellite platform and the payloads of \emph{Insight-}HXMT. The coordinate system used in this study is shown on the upper-left. It is adapted from Fig.~1 in \cite{LUO20201}.
  }\label{HEjpg}
\end{figure*}

The High Energy X-ray Telescope~\citep[HE,][]{HEofHXMT} consists of the main detector (HED),
high energy collimator (HEC), auto-gain control detector
(HGC), anti-coincidence shield detector (HVT), particle
monitor (HPM), data processing and control box (HEB) and
power box (HEA). As the main detector of HE, HED is responsible for the
observation of celestial sources. It consists of 18 NaI(Tl)/CsI(Na) phoswich detectors (labelled as HED-0, 1 ... to 17), each with a diameter of 190 mm and a collimator made of lead and tantalum. The NaI detectors are sensitive to the hard X-rays in 20-250~keV, whereas the CsI acts as an anti-coincidence detector to reduce the background of NaI. The thickness of the NaI and CsI crystals are 3.5 mm and 40 mm, respectively. HGC provides the
auto-gain control and energy calibration for HED. HVTs act
as an active shielding system to reduce HED's background
caused by charged particles. HPMs monitor the flux of
charged particles and send out an alert to switch off the high
voltages of HEDs and HVTs in high flux regions to avoid
potential damage to the PMTs of these detectors. The overview of all telescopes aboard \emph{Insight-}HXMT is shown in Figure~\ref{HEjpg}.

\emph{Insight-}HXMT/HE can observe the high energy sky with two observation scenarios~\citep[][]{OverviewHXMT} which can work simultaneously:
\begin{itemize}
\item[(1)] Collimator-based (pointed or scan) observation: using NaI/CsI detectors, HE can detect hard X-ray emission in 20-250~keV from the celestial sources within the FOV defined by collimators. HE can do pointed observation to a specific source, or scan observation to a small sky area.
\item[(2)] Gamma-ray all-sky monitoring: Gamma-ray photons with energy higher than $\sim$200~keV from all directions can penetrate the spacecraft and be detected by the CsI detectors. CsI can monitor the gamma-ray all-sky with the only exception of the occulted region by the Earth. This gamma-ray all-sky monitoring is an extended capability for HE and inapplicable for ME or LE telescopes aboard \emph{Insight-}HXMT.
\end{itemize}

The primary observation scenario of HE is the collimator-based observation consisting of pointed observation and scan observation, both of which use the NaI as the main detector and detect sources within the field of view (FOV) of collimator in 20-250 keV. The CsI detectors mainly act as the anti-coincidence and shielding to reduce the background of NaI. Due to the high thickness of the CsI(Na) crystals, high energy (greater than about 200 keV) gamma-ray photons can penetrate the spacecraft and be recorded by the CsI detectors, thus the GRBs unocculted by the Earth can be detected. For each NaI/CsI phoswich detector of HE, the event-by-event data (called EVT data hereafter) is recorded and we can use the pulse
width information in the EVT data to distinguish between NaI events and CsI events. For this paper, we only select these events recorded by CsI detectors by screening events with pulse width greater than 75. The response and calibration of the CsI detectors are mainly described in Section~\ref{subsec:rspintro}.
\label{sec:intru}
The gain of NaI/CsI detector is mainly controlled by the High Voltage of the PMT which converts the scintillation light produced by NaI/CsI crystals to the electric signals. To meet various observation requirements, two gain modes are designed for the HE NaI/CsI detectors: normal gain (NG) mode and low gain (LG) mode. The measured energy ranges for CsI detector are listed as bellow: the NG mode in $40-800$~keV and the LG mode in 200-3000~keV (both refer to the deposited energies).
NG mode is the main working mode that the auto-gain control system can keep the full-energy-peak of the 59.5 keV photons (emitted from a radioactive source $^{241}$Am) in a fixed channel of the NaI detector. In LG mode, the high voltage of the each \emph{Insight-}HXMT/HE detector is reduced and the auto-gain control system is disabled to achieve a higher energy range detection. As derived from the two-year observations since \emph{Insight-}HXMT has operated in-orbit, the E-C relationships of each \emph{Insight-}HXMT/CsI detector for both NG and LG modes vary over time, so calibration of the instrumental response is of great importance.

\section{Search for Bursts} \label{sec:algo}
A pipeline is developed to search for GRBs
for blind search and the coherent search method for targeted search~\citep{10.1093/mnras/stab2760}. The blind search software continually monitors each group (18 CsI detectors are divided into 6 groups, 3 in each group) of detectors count rates, for which the reason is to reduce statistical fluctuations of one or a few detectors, one of the main causes of false trigger. The pipeline is triggered when counts rates in the EVT data of three or more groups of CsI detectors exceed the background counts rates, by 3$\sigma$ on five timescales:  0.05s, 0.1s, 0.2s, 0.5s and 1s by calculating an average of the counts rates of ~10 s of the previous data as background.

If a GRB detected by other external missions, falls in the field of view of HE, but did not trigger
the bind search, targeted search in HE/CsI data would be launched. The targeted coherent search method is applied to the EVT data of HE CsI detectors, which can recover true astrophysical bursts that are too weak to trigger through blind search. This coherent method was originally developed by \cite{Blackburn_2015}, improved by \cite{2016arXiv161202395G}, \cite{2019arXiv190312597G}, and verified by \cite{Kocevski_2018}. From the statistics of searched GRBs, we find the fluence (from GBM Burst Catalog) of HE targeted search could reach $\sim 10^{-7}$~$\rm erg$ $\rm cm^{-2}$ (10
keV--1 MeV) in duration of 0.256~s. For this paper, 15 GRBs that are found from targeted search are included and labeled in Table~\ref{tab:main_table}, while the rest are from blind search. Details of burst search are presented in a separate paper~\citep{10.1093/mnras/stab2760}.

\section{Catalog Analysis } \label{sec:catalogana}
\subsection{Instrument Response} \label{subsec:rspintro}
There are two parts in the instrumental response of CsI detectors: the energy redistribution of the photons from incident energy to deposition energy that is determined by the property of the CsI crystal and the mass distribution of satellite; and the Energy-Channel (E-C) relationship that is determined by the detectors and electronic system of the instrument. For the photon energy redistribution, the reliability is mainly determined by the accuracy of the mass model of the satellite and the payloads~\citep{Agostinelli2003G}. Based on the initial mass model of the satellite platform~\citep{Xie2015},
we calibrate the mass model with the Crab pulse radiation as a standard candle~\citep{Li2018},
to generate accurate instrumental response for the all sky gamma-ray monitoring.

The in-flight E-C relationship and energy resolution of the instrument can be obtained by analyzing
the characteristic lines of the in-orbit observed and the on-ground simulated background spectra~\citep{groundofHE}.
The E-C relationship varies over time, whereas the energy resolution remains stable~\citep{LI202064}.
Therefore, we update the E-C relationship every month, and take the average energy resolution of all calibration results to generate the instrumental response.

After the above calibration, a new response matrix library is established and a simulated spectral analysis is performed to test the HXMT/CsI
spectroscopy capabilities. Calibration of the instrumental response to GRB is carried out
with GRBs \citep{LUO20201}. In a GRB observation, the incident direction of the GRB photons is supposed to be arbitrary, however, only the instrumental response to several directions can be calibrated directly.
A common method of the instrumental response testing is the cross-calibration with other instruments by comparing the energy spectrum of the simultaneously observed GRB~\citep{Sakamoto2011,Tsujimoto2011,Tierney2011The,Ishida2014}. The detection efficiency of HXMT/CsI is checked by the joint spectral analyses with \emph{Fermi}/GBM,
\emph{Swift}/BAT and Konus-\emph{Wind}, in which we find that HXMT/CsI can provide better constraint on GRB spectrum at higher energy band~\citep{LUO20201}. The instrumental responses are mainly
obtained by Monte Carlo simulation with the Geant4 tool and the mass model of both the satellite and all the payloads, which is
updated and tested with the Crab pulse emission in various incident directions. Both the Energy-Channel relationship and the energy resolution are
calibrated in two working modes (NG mode and LG mode) with the different detection energy ranges.
The simulated spectral analyses show that HXMT/CsI can constrain the spectral parameters much better in the high energy band than that in the low energy band.
As introduced in \citep{LUO20201}, the joint spectral analyses are performed to ten bright GRBs observed simultaneously with HXMT/CsI and other instruments (\emph{Fermi}/GBM, \emph{Swift}/BAT, Konus-\emph{Wind}), and the results show that the GRB flux given by HXMT/CsI is systematically higher by $7.0\pm8.8\%$ than those
given by the other instruments. The HXMT/CsI-\emph{Fermi}/GBM joint fittings also show that the high energy spectral parameter can be constrained much better as the HXMT/CsI data are used in the joint fittings.

\begin{figure*}
\begin{center}
\includegraphics[width=0.8\textwidth]{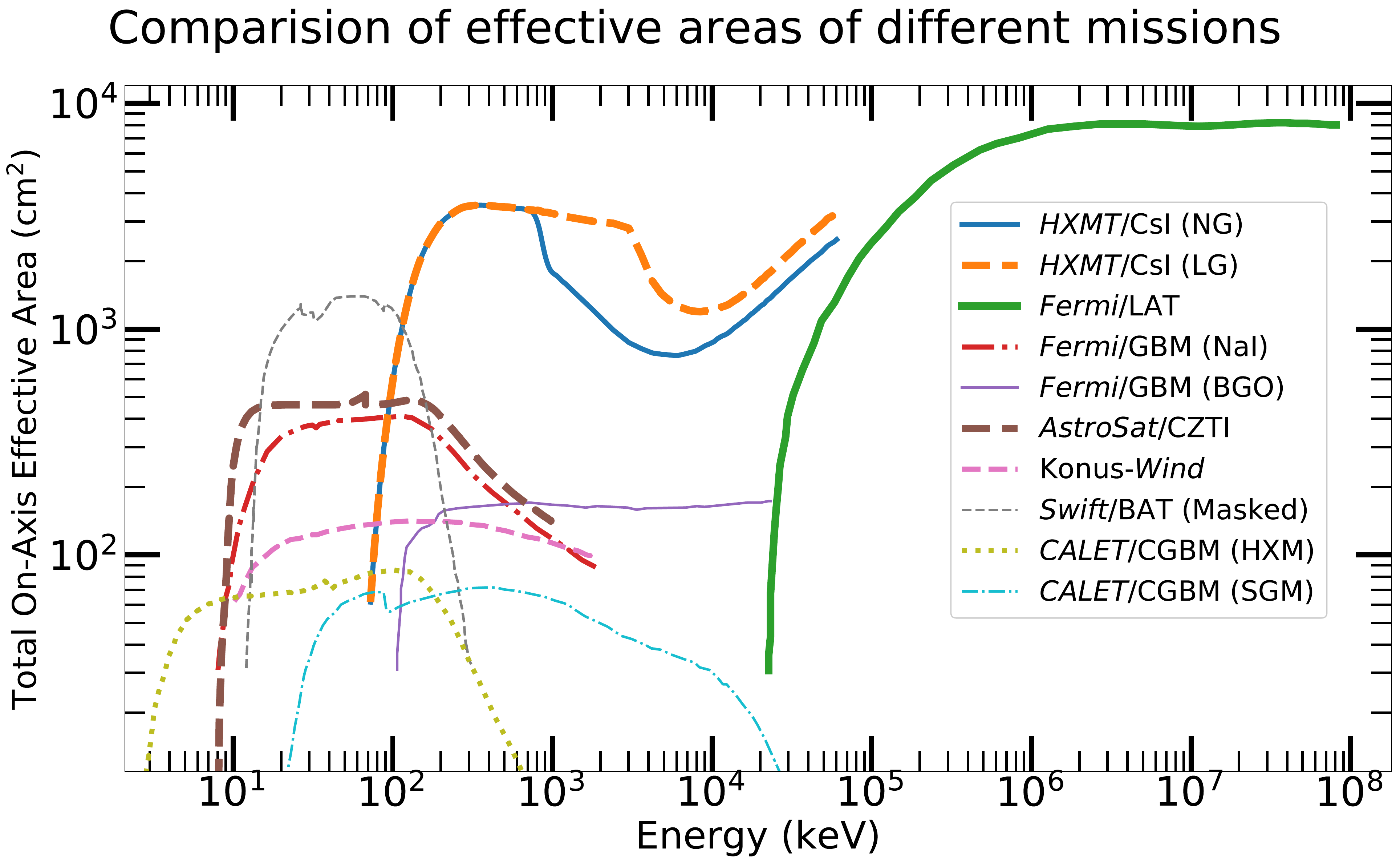}
\caption{\label{fig:ARF_allsatjpg} Effective areas of HXMT/CsI, \emph{Fermi}/LAT, \emph{Fermi}/GBM, Konus-\emph{Wind}, \emph{Swift}/BAT, \emph{CALET}/CGBM and \emph{AstroSat}/CZTI. The effective area of \emph{Fermi}/GBM (NaI) is the averaged over the unocculted sky. It is adapted from Fig.~13 in \cite{LUO20201}. }
\end{center}
\end{figure*}

In summary, the instrumental responses of the HXMT/CsI detectors of \emph{Insight-}HXMT are well calibrated in aspects of mass model, E-C relationship and the energy resolutions for both the NG and LG modes. Thanks to the large effective area in the high-energy band, as shown in Figure~\ref{fig:ARF_allsatjpg}, HXMT/CsI shows its advantages in constraining the GRB spectra in high energy band, together with other missions (such as \emph{Fermi}/GBM) which provide observations at lower energies.

\begin{figure*}
\begin{center}
 \centering
  \includegraphics[width=0.8\textwidth]{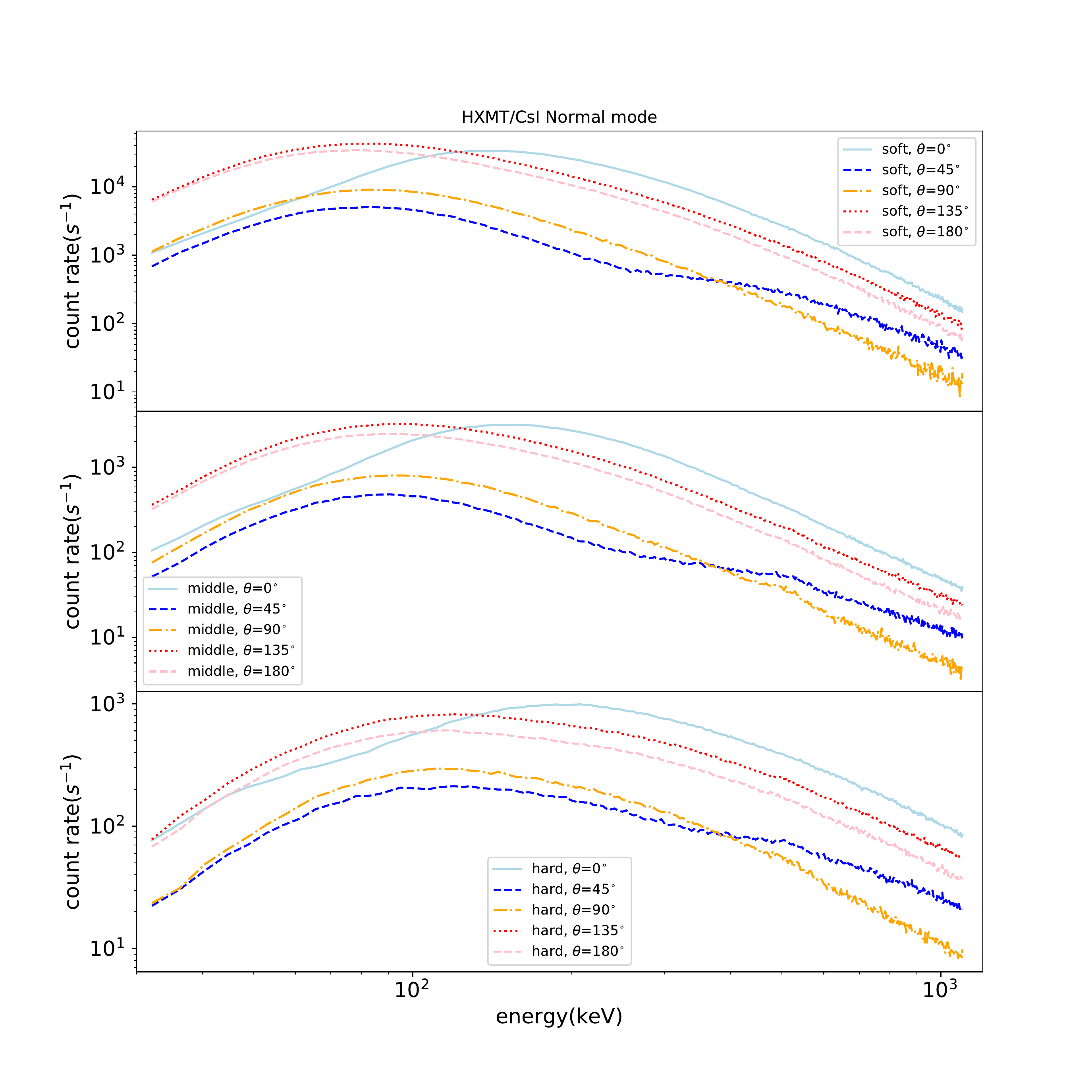}\\
\caption{\label{fig:Norspec} The deposited energy spectra of three types of BAND with incident angle $\theta$ of $0^{\circ}$ to $180^{\circ}$ are shown in NG mode. The azimuthal angle $\phi$ is $0^{\circ}$. Here the incident angle $\theta$ and the azimuthal angle $\phi$ denote the direction of gamma rays in spherical coordinates of HE/CsI detector.}
\end{center}
\end{figure*}

\begin{figure*}
\begin{center}
 \centering
  \includegraphics[width=0.8\textwidth]{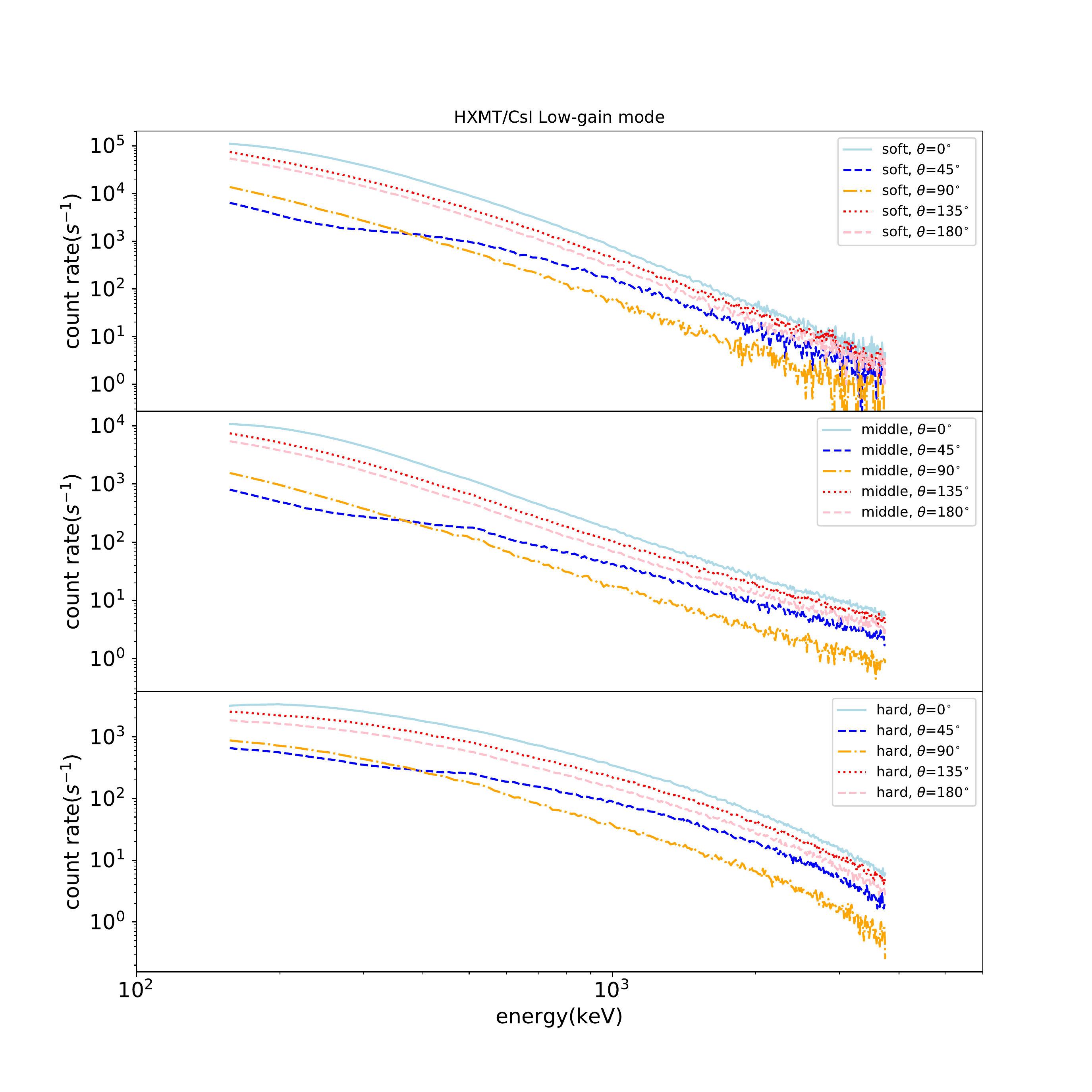}\\
\caption{\label{fig:LGspec} The deposited energy spectra of three types of BAND with incident angle $\theta$ of $0^{\circ}$ to $180^{\circ}$ are shown in LG mode. The azimuthal angle $\phi$ is $0^{\circ}$. Here the incident angle $\theta$ and the azimuthal angle $\phi$ denote the direction of gamma rays in spherical coordinates of HE/CsI detector.}
\end{center}
\end{figure*}

Note that the the deposited energy range is usually very different from incident energy bands due to the responses of HXMT/CsI detectors. Three types of spectra are listed in Table~\ref{tab:softmidhardspectra}, where the soft and middle spectra are with BAND model, the hard spectra is with CPL model, and they are representative of GRBs of different hardness; details of BAND and CPL model are clarified in Section~\ref{subsubsec:TAfit}. They are simulated with the same amplitude in each type of spectra (use $fakeit$ in Xspec 12.11.0, with all the model normalization factors equal to 1), and convoluted with the total response of 18 HXMT/CsI detectors. The deposited energy spectra of incident angle theta from $0^{\circ}$ to $180^{\circ}$ are shown in NG mode (Figure~\ref{fig:Norspec}) and in LG mode (Figure~\ref{fig:LGspec}) respectively, which shows the ranges of deposited energy. Considering the deposited spectra and effective area shown in Figure~\ref{fig:ARF_allsatjpg}, the deposited energy range of HXMT/CsI data is determined to be about 150-800~keV for NG mode and about 200-3000~keV for LG mode in the following spectra analysis.
\begin{table}[]
\centering
\caption{\label{tab:softmidhardspectra} \textbf{Three types of spectra and parameters. The  soft and middle spectra are with BAND model, and the hard spectra is with CPL model.} }
\begin{tabular}{llll}
\hline
Types of spectra  &$\alpha$ &$\beta$ &$E_{\rm peak}$(keV)   \\
\hline
 SOFT &-1.9 &-3.7 &70 \\
 MIDDLE  &-1.0 &-2.0 &230 \\
HARD &-0.5 &-  &1500 \\
\hline
\end{tabular}
\end{table}

\subsection{GRB Samples}\label{subsec:detecsel}
In summary, we define three types of GRB samples as shown in Table~\ref{tab:GRBsample}. Only the `GOLDEN' GRBs could be utilized in the joint analysis, which are required to be located well and detected without saturation in HXMT/CsI data. The `SILVER' samples denote those which are detected by other external missions, but not located or lack of data for joint analysis. In this analysis, public data offered by \emph{Fermi}/GBM, \emph{Swift}/BAT or GECAM are utilized for joint analysis. The bursts which are only detected by HXMT are labeled as `BRONZE'. \textbf{The HXMT/HE data files will not be used if there is saturation in electronics subsystem during the GRBs, or the data quality is not good (such as data loss). Saturation is defined when the two following criteria are matched simultaneously: firstly, total count rate of 18 CsI detectors exceeds 20000 counts/s, as evaluated over four different timescales: 0.1, 0.2, 0.5, and 1.0 s; secondly, in at least one of the three groups of 6 CsI detectors that share the same Analog-to-Digital Converter (ADC)~\citep[][]{HEofHXMT}, the total light curve has at least one 5-ms bin with zero counts.} As shown in Table~\ref{tab:GRBsample}, this kind of GRB samples are labeled as `IRON', which are not utilized in the following analysis.

\begin{table*}[]
\centering
\caption{\label{tab:GRBsample} Classification of four kinds of \emph{Insight-}HXMT GRB samples.}
\begin{tabular}{llllll}
\hline
Type  &Location and data for joint analysis  &External detected &Saturation in HXMT data & Utilized or  not & Number \\
\hline
GOLDEN &Yes &Yes &No &Yes &202 \\
SILVER &No &Yes &No &No &44 \\
BRONZE &No &No  &No &No &50 \\
IRON  &Yes/No &Yes/No &Yes &No &26\\
\hline
\end{tabular}
\end{table*}

\subsection{Suppression of Background From Charged Particles}
Among the main detectors of the HE telescope, HVTs act as an active shielding system to reduce HED background caused by charged particles ~\citep[HE,][]{HEofHXMT}. In order to suppress the background from charged particles, events which are coincidence with signal in HVTs are removed since they are likely produced by charged particles rather than gamma-rays. \textbf{Besides, those events which also have energy deposition in NaI (Tl) detectors are also eliminated, because the signals of NaI (Tl) are mainly from the target source when we perform pointed or scan observation.} Response of 18 CsI detectors in HED according to Section~\ref{sec:instru} are different but the discrepancy between them are small (unlike the case in \emph{Fermi}/GBM or GECAM where detectors have different orientations), therefore, all dat events in 18 CsI detectors are utilized in this study.

\subsection{Dead Time Correction}\label{subsec:deadcor}
\begin{figure*}
\begin{center}
 \centering
  \includegraphics[width=0.8\textwidth]{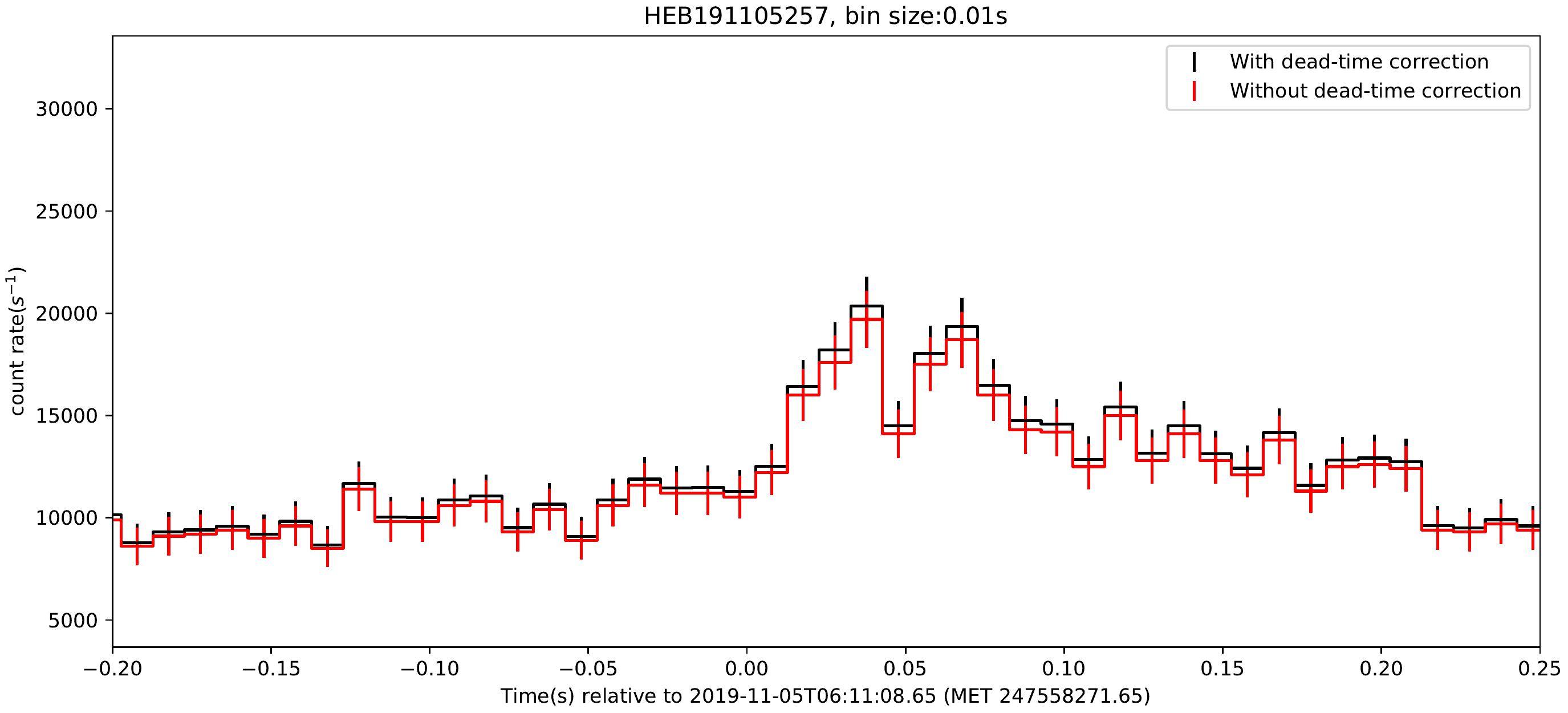}\\
\caption{\label{fig:DTcompare} A comparison between light curves with and without dead-time correction of a short GRB triggered at 2019-11-05T06:11:08.65 UTC. The histogram of red lines and error bars denote the light curve without dead-time correction, while the black with dead-time correction. Here MET means the mission elapsed time of HXMT.}
\end{center}
\end{figure*}

 Dead time of HED detectors is complicated. According to \cite{XIAO202058}, every 6 detectors in 18 NaI (Tl)/CsI (Na) detectors in HE share one ADC read-out electronics, thus a signal in any detector will cause dead time not only to the incident detector itself, but also to the other 5 detectors of the same ADC. The dead time of one event is normally about $4\sim8$~$\mu$s, but it will be longer for the signals generated by high
energy charged particles. There is a dead time counter in HEB for each HED, and the information is recorded in the form of proportion of dead time. The proportion of dead time in a time interval is determined with the method established in \cite{XIAO202058}. The correction factors on the light curves are determined by $\frac{1}{1-P_{\rm dt}}$, where $P_{\rm dt}$ is the proportion of dead time in every time bin with bin width of 10 ms. The light curves of each CsI detector are corrected by multiplying with series of dead-time correction factors before background fitting is performed. As shown in Figure~\ref{fig:DTcompare}, the comparison between light curves with and without dead-time correction of a short GRB is present, which shows that the dead-time correction is necessary for refined analysis.

\subsection{Background Fitting}
After the data have been selected for a given GRB, a background model in the form of polynomial function with time, is computed separately for each detector, based on user-selected time intervals for background region. Here we use a pre-burst interval and a post-burst interval as user-selected background regions. The order of polynomial is determined by the best goodness of fits with polynomials of different orders, where the order could be 0, 1, 2 respectively. For each detector, background fitting is carried out for each energy channel, and the background fit determined from pre-burst and post-burst intervals are used to estimate the background counts in the signal region of the burst. The net counts number of GRB in each channel is obtained by subtracting the counts of background from the total counts number. Background fitting is performed in each detector channel by channel in deposited energy spectrum, until the net counts and background counts of all channels for the given GRB are determined. Statistical errors of net counts are determined by the root of the sum of squares of the statistical errors of background estimation and Poisson errors of total counts (i.e. classical error propagation). Note that all light curves for each energy channel in each detector have been corrected with dead-time correction factors. The source time intervals for all detectors are chosen to be same for the GRB data analysis, thus data of all detectors could be added up to reduce the statistical uncertainties.

\subsection{Source Time Interval Selection}
 Source time interval selection may be different when we perform different analyses. For the computation of duration, peak fluxes and fluence, the source interval is loosely selected, by starting from several seconds before the burst begins and ending at several seconds after the burst ends, and may have overlaps with pre-burst and post-burst intervals for background selection. If there exists difference of start- and end-time of a GRB between HXMT/CsI and other missions due to the energy detection range, the loosest source interval is applied for the joint data analysis of these missions. For the analysis of time-average spectra, the source time interval is usually selected to be narrower. The events in $T_{\rm 90}$ or the whole duration in some cases based on deposited photon counts are selected for the time-averaged energy spectra.

\begin{figure}
\begin{center}
 \centering
  \includegraphics[width=\columnwidth]{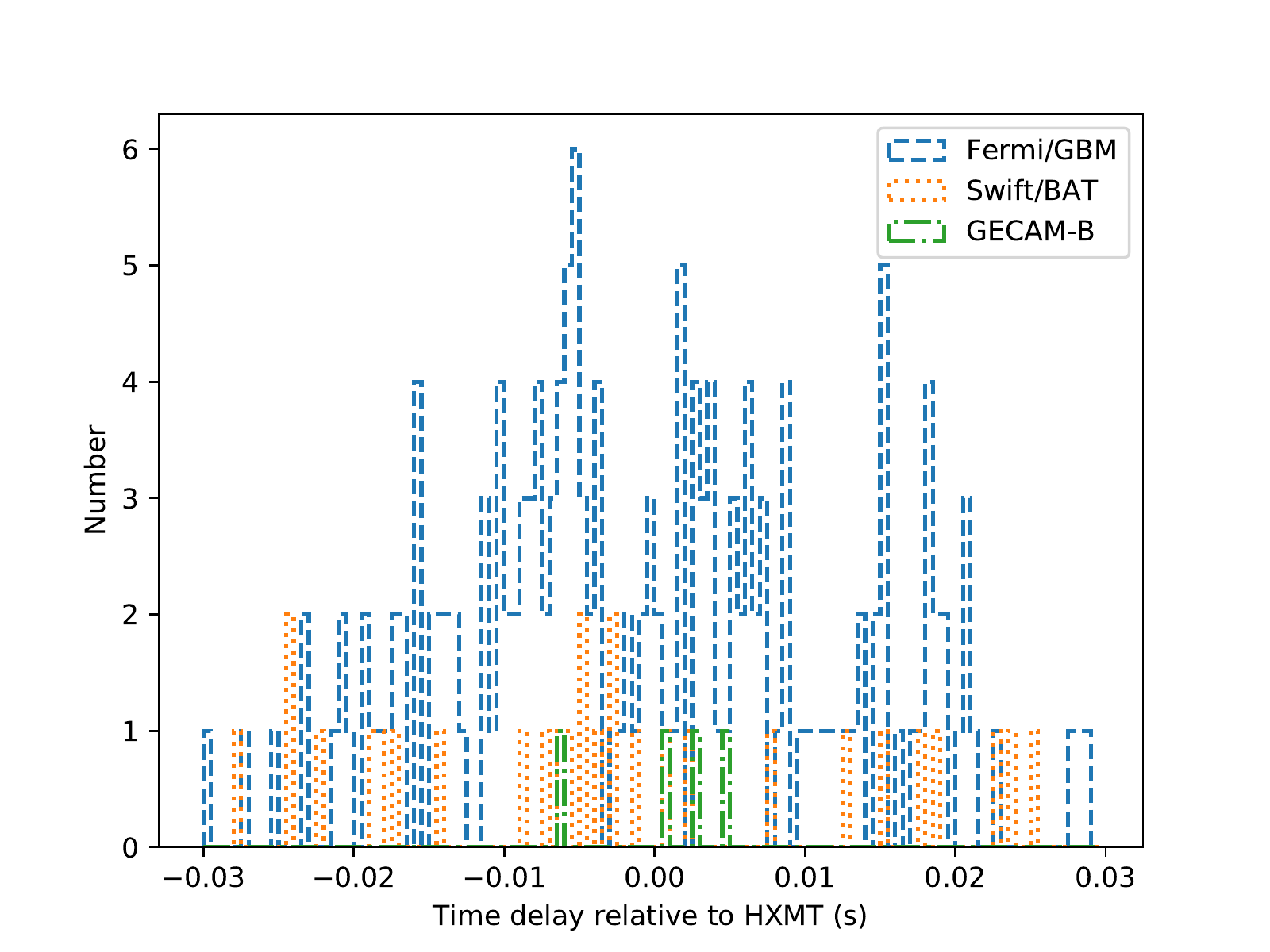}\\
\caption{\label{fig:timedelay} The distribution from `GOLDEN' GRB samples of time delays of \emph{Fermi}/GBM~(dashed), \emph{Swift}/BAT~(dotted), and GECAM~(dot-dashed) relative to HXMT are shown.}
\end{center}
\end{figure}

 In the case of joint analyses with other missions, the time delay due to the different arrival times of two spacecrafts is considered. The method mentioned in \cite{Hurley_1999} is utilized to determine the time delay between HXMT and other missions. Figure~\ref{fig:timedelay} shows the time delays of other missions relative to HXMT for `GOLDEN' GRB samples. The red histograms denote the time delays of \emph{Fermi}/GBM, while the blue ones denote those of \emph{Swift}/BAT and the green ones for those of GECAM. The negative values represent that GRBs arrived at other missions earlier than HXMT, while the positive numbers means that the GRB arrival time of HXMT is earlier. Some time delays reach up to tens of milliseconds and must be considered, because the time scale (64~ms) of deconvolved photon spectra is in the same order of magnitude.


\subsection{Duration, Fluence and Peak Flux} \label{subsec:DPF}
In order to take advantages of different missions and to reduce uncertainties in determination of the parameters of GRBs, joint spectral analyses between HXMT and \emph{Fermi}/GBM, \emph{Swift}/BAT and GECAM are applied on all `GOLDEN' GRBs. This paper reports various measures of the duration, peak fluxes and fluence of each burst.
\textbf{We take GRB 200125B (HEB200125863), GRB 210121A (HEB2101210779) and GRB 210112A (HEB210112068) as samples to show the procedures performed on all GRBs in Appendix A.} The fits to time-averaged and time-resolved spectra are all performed with Xspec 12.11.0.

\subsubsection{Time-Average Spectrum}\label{subsubsec:TAfit}
The time-averaged energy spectra are obtained from HXMT/CsI together with other instruments. For example, from \emph{Fermi}/GBM, the spectra from two NaI(Tl) scintillation detectors working at 8~keV $-$ 1~MeV and one BGO scintillation detectors at $\sim200$~keV to $\sim4$~MeV are selected according to the statistics of GRB signal. A joint fit with time-averaged energy spectra from HXMT/CsI and three (two NaI and one BGO) spectra from \emph{Fermi}/GBM are performed, where the parameters of GRB spectral shape are shared but the amplitudes are set to be float for these four sets of data, to compensate the possible lack information either in the adopted spectrum model or the understanding of the instruments response. In most cases, the amplitudes of fitted spectra are consistent well with each other within one standard deviation.

To perform the joint fit, 6 models are chosen including a single power law (PL), Band's GRB function (BAND,~\cite{1993ApJ...413..281B}), an exponential cut-off power law (CPL, also called Comptonized Model) and plus a component of black body function (BB) of float amplitudes.
The PL model is represented with Equation~\ref{func:PL} with two free parameters, where A denotes amplitude and $\alpha$ is the spectral index.

\begin{equation}\label{func:PL}
N_{\rm PL}(E)=A E^{\alpha}.
\end{equation}
The BAND function has four free parameters: low and high energy spectral indices, denoted as $\alpha$ and $\beta$ respectively, the peak energy of $\nu F_{\nu}$ spectrum, denoted as $E_{peak}$, and amplitude, as shown in Equation~\ref{func:BAND}.
\begin{equation}\label{func:BAND}
\begin{split}
&  N_{\rm BAND }(E) = \\
& A \begin{cases}
\biggl(\frac{E}{100 \ \rm keV }\biggr)^{\alpha} \exp \biggl[- \frac{ (\alpha +2) E}{ E_{\rm peak} } \biggr], \ E \geq \frac{ (\alpha - \beta) \
E_{\rm peak} } { \alpha +2} \\
\biggl( \frac{E}{ 100 \ \rm keV } \biggr)^{ \beta } \exp (\beta -\alpha) \biggl[ \frac{(\alpha-\beta ) E_{\rm peak}}{100 \ \rm keV \ (\alpha
+2)} \biggr]^{\alpha-\beta }, \\ E < \frac{(\alpha -\beta ) \ E_{\rm peak}}{\alpha +2}.
\end{cases}
\end{split}
\end{equation}

CPL model is a subset of BAND model if $\beta$ is very small and the part of $E < \frac{(\alpha -\beta ) \ E_{\rm peak}}{\alpha +2}$ of BAND model is ignored. There are three parameters in CPL model: the amplitude A, the lower energy index $\alpha$, and the $\nu F_{\nu}$ peak energy, $E_{\rm peak}$, as shown in Equation~\ref{func:CPL}.
\begin{equation}\label{func:CPL}
N_{\rm COMP}(E) = A \ \Bigl(\frac{E}{E_{\rm piv}}\Bigr) ^{\alpha} \exp \Biggl[ -\frac{(\alpha+2) \ E}{E_{\rm peak}}  \Biggr].
\end{equation}

The Equation~\ref{func:BB} for the photon spectrum of a black body usually used in the spectral fitting, where $K = L_{\rm 39}/D^2_{\rm L,10kpc}$ is defined by the blackbody luminosity L in units of $10^{39}$ erg s$^{-1}$ in the GRB host galaxy frame and the luminosity distance $D_{\rm L}$ in units of 10 kpc.
\begin{equation}\label{func:BB}
N_{\rm BB}(E) = \frac{K \times 8.0525 E^{2} }{(kT)^{4}[\exp (E/kT) -1]}.
\end{equation}
 All models are formulated in units of photon flux with energy (E) in keV and multiplied by a normalization constant A (ph s$^{-1}$ cm$^{-2}$ keV$^{-1}$), and $kT$ and $E$ are measured in the observer's frame. We determine the best spectral parameters by optimizing the Castor C statistic value. Castor Cstat \citep[henceforth Cstat,][]{cash} is a likelihood technique modified for a particular data set to converge to a $\chi^2$ with an increase of the signal. $\chi^{2}$ is evaluated for each spectral fit that is performed through minimizing C-Stat. \textbf{Among these 6 models, the best model is determined when a
single additional parameter changes in $\chi^{2}$ by
at least 6 since the probability for achieving this
difference is 0.01 as suggested by~\cite{Goldstein_2012}, which is a conservative threshold to avoid the false positive rates of the extra component.}

As shown in Figure~\ref{spectrum_GRB200125B} in Appendix A, for GRB 200125B (HEB200125863), joint spectra fit with BAND model is the best model, with a reduced $\chi^2$ of 1.14. A BAND model of low energy spectral index $\alpha=-0.78\pm0.01$, high energy spectral index $\beta=-2.70\pm0.03$, and the $\nu F_{\nu}$ peak energy, $E_{peak}=200.8\pm7.7$~keV, is determined. GRB 210121A is observed by HXMT/CsI, \emph{Fermi}/GBM and GECAM, and joint analysis with the other two missions is shown in Figure~\ref{spectrum_GRB210121A}. \textbf{$\alpha=-0.64\pm0.01$, $\beta=-2.49\pm0.03$ and  $E_{peak}=932.4\pm23.8$~keV are determined with a BAND model.} Figure~\ref{spectrum_GRB210112A} shows the spectrum of GRB 210112A detected by \emph{Swift}/BAT and HXMT/CsI. The best model is determined to be CPL, with $\alpha=-1.27\pm0.03$, and $E_{peak}= 347.7 \pm 49.7$~keV for the CPL component. In the case of utilizing data of \emph{Swift}/BAT only, PL model is determined and the $E_{peak}$ is not measured due to lack of information from higher energy band.

\begin{figure}[!htbp]
  \centering
  \includegraphics[width=\columnwidth]{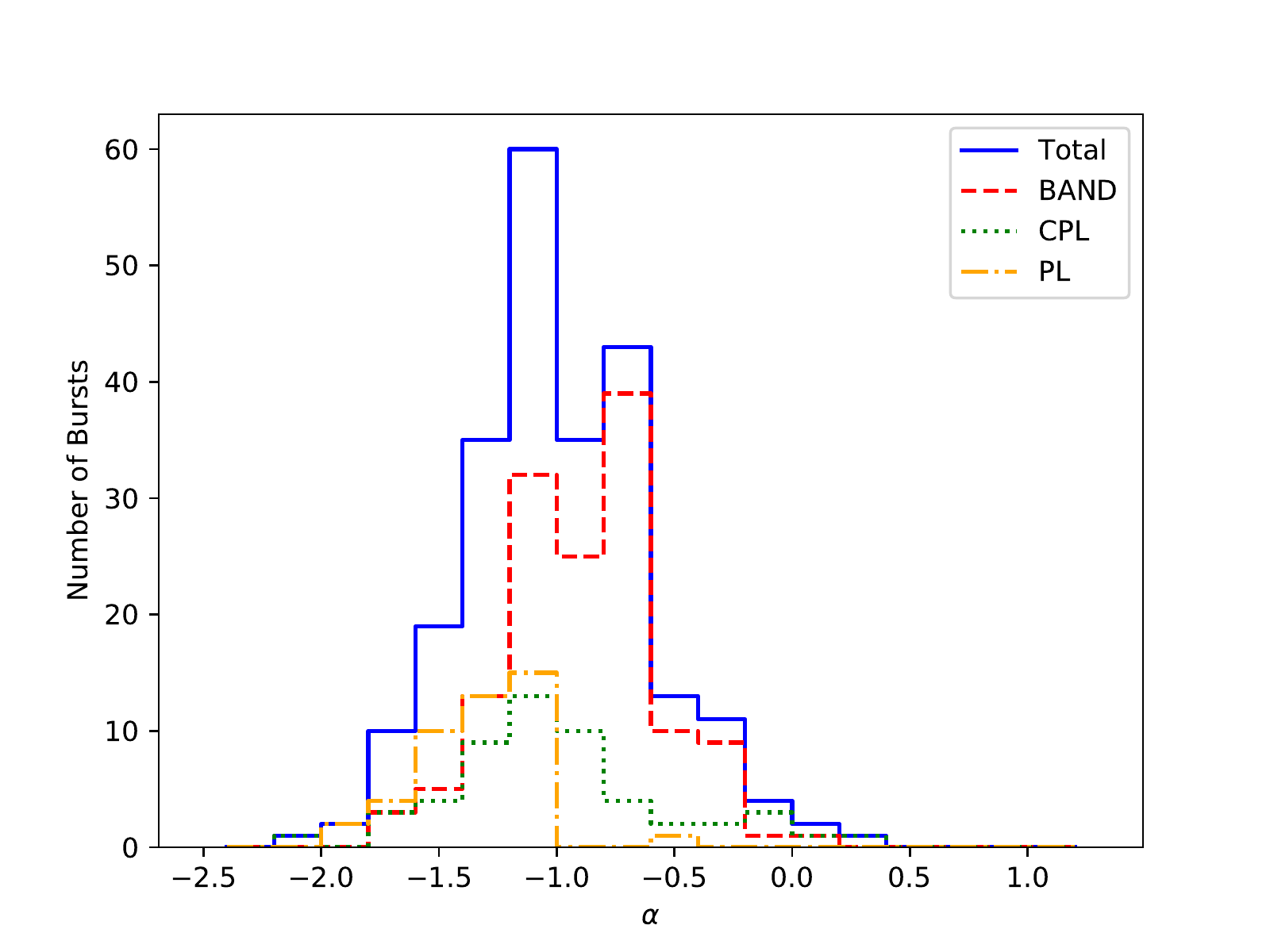}\\
  \caption{The distribution of low energy spectral index $\alpha$ of `GOLDEN' GRBs from time-averaged spectral fits are shown. The blue solid lines denote the total distribution of $\alpha$ from BAND model (red dashed lines), CPL model (green dotted lines) and PL (orange dot-dashed lines).
  }\label{Fig:dis_alpha}
\end{figure}

\begin{figure}[!htbp]
  \centering
  \includegraphics[width=\columnwidth]{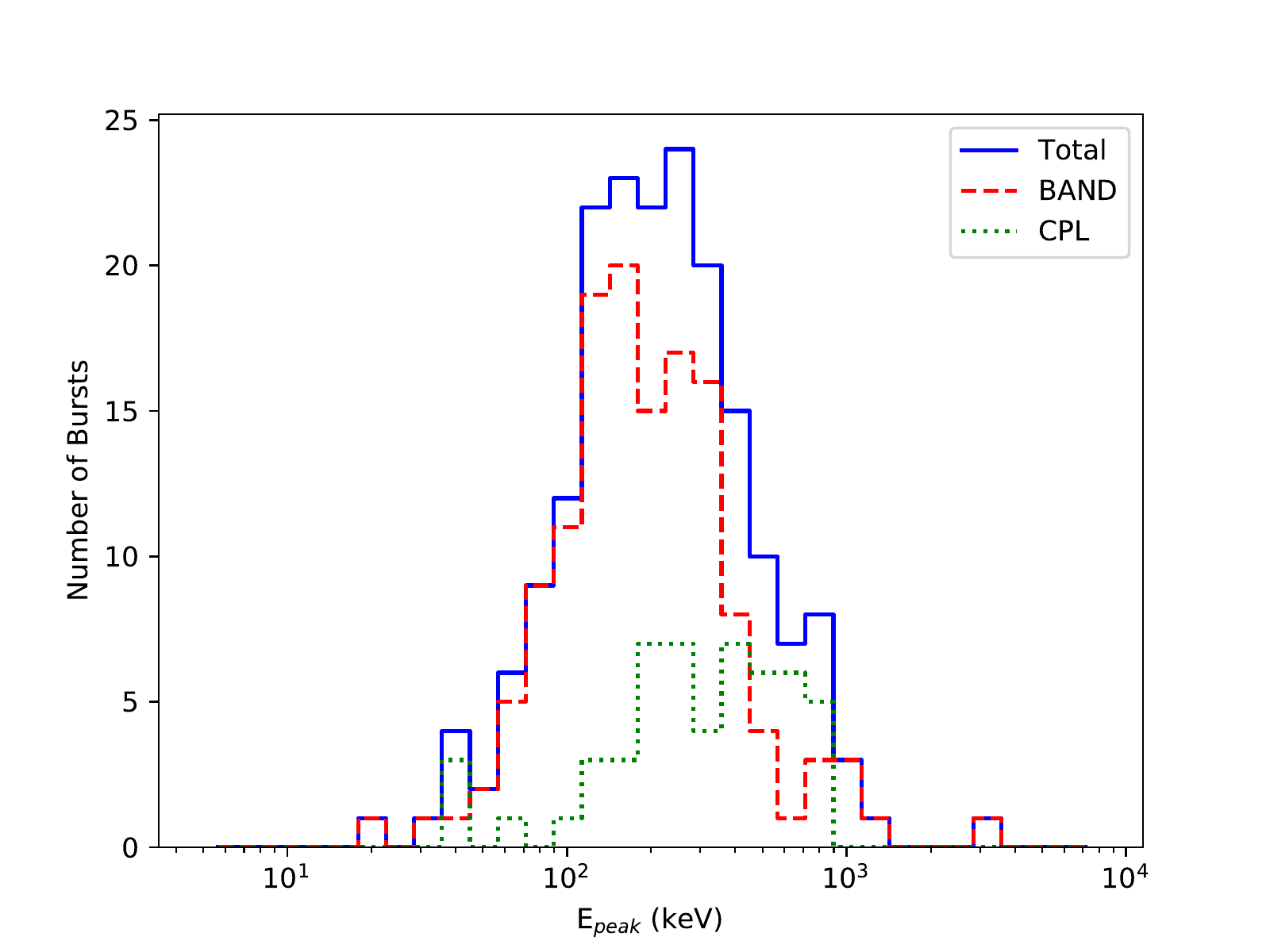}\\
  \caption{The distribution of $\nu F_{\nu}$ peak energy $E_{peak}$ of `GOLDEN' GRBs from time-averaged spectral fits are shown. The blue solid lines denote the total distribution of $E_{peak}$ from BAND model (red dashed lines), CPL model (green dotted lines).
  }\label{Fig:dis_Epeak}
\end{figure}

\begin{figure}[!htbp]
  \centering
  \includegraphics[width=\columnwidth]{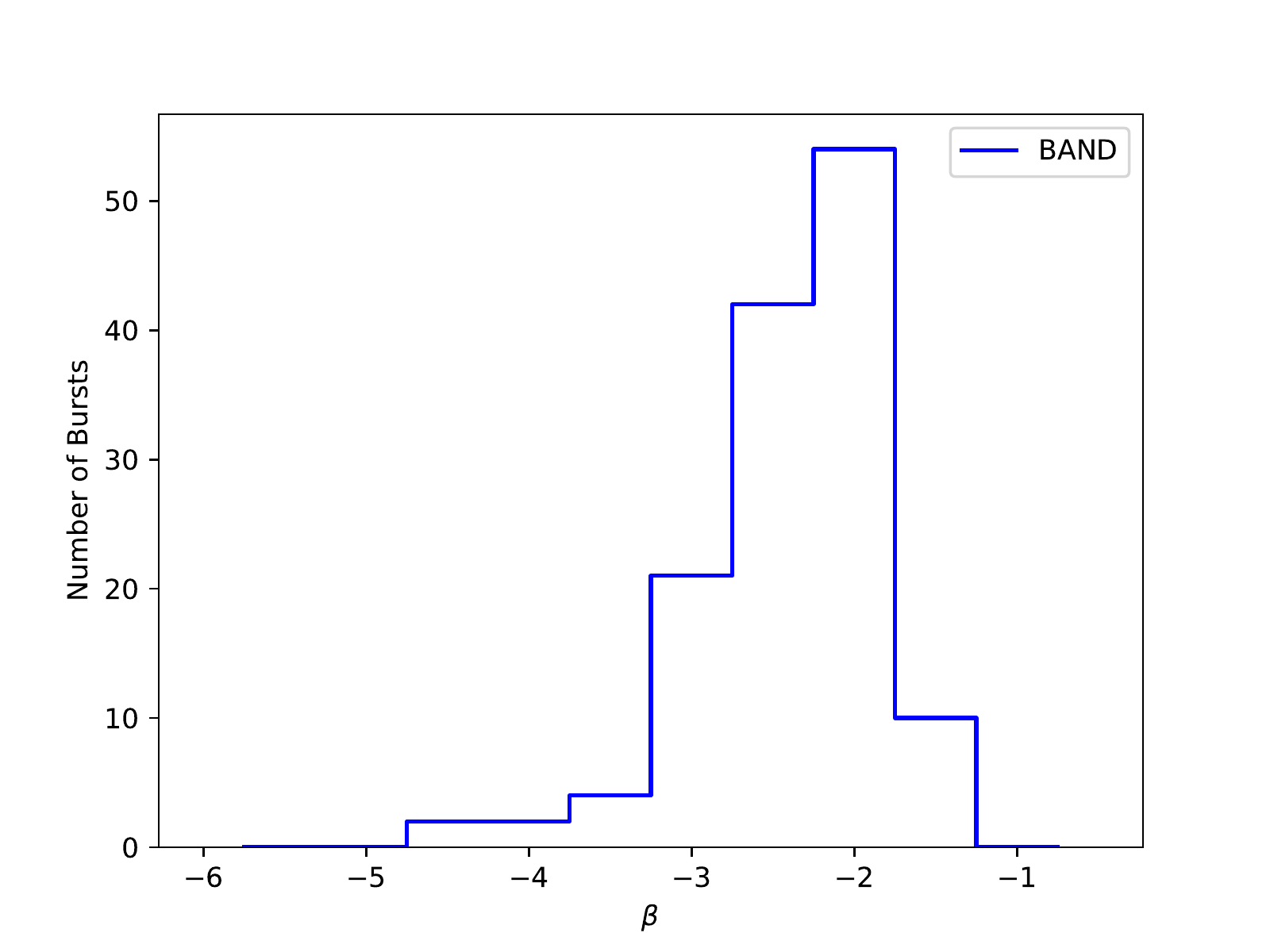}\\
  \caption{The distribution of high energy spectral index $\beta$ of `GOLDEN' GRBs from time-averaged spectral fits are shown.
  }\label{Fig:dis_beta}
\end{figure}

The distribution of energy spectra parameter of `GOLDEN' GRBs are shown in Figures~\ref{Fig:dis_alpha}, ~\ref{Fig:dis_Epeak}, ~\ref{Fig:dis_beta}. Up to 29\% low energy spectral indices of the `GOLDEN' samples violate -2/3 synchrotron ``line-of-death"~\citep{Preece_1998}, while an additional 6\% of the indices violate the -3/2 synchrotron cooling limit. The distribution of high-energy indices in Figure~\ref{Fig:dis_beta} peaks at a slope slightly steeper than -2 and have a extension toward steeper values. The $E_{peak}$ distribution generally peaks around 200~keV, and spans over two orders of magnitude.


\subsubsection{Duration and Fluence}\label{subsubsec:DurationandFluence}
For `SILVER' and `BRONZE' samples, the burst durations are determined by integrating the signal counts from GRBs with deposited energy in ranges of 40-800~keV in NG mode and 200-3000~keV in LG mode, respectively. The duration $T_{\rm 50}$ and $ T_{\rm 90}$ are determined from the interval between the times where the burst has reached 25\% (5\%) and 75\% (95\%) of its maximum counts. The method developed for BATSE \citep{koshut1996} are used to determine the systematic error. Table~\ref{tab:silbrodurations} shows the values of durations of `SILVER' and `BRONZE' samples.

For `GOLDEN' samples, the burst durations $T_{\rm 50}$ and $ T_{\rm 90}$ are computed in the 10-2000~keV energy range. They are determined using a method similar to that developed for BATSE~\citep{koshut1996}, the counts spectra of HXMT/CsI and the other missions in each time bin 64 ms are deconvolved and the durations are computed from the time history of fitted photon spectra, as shown in Figure~\ref{flux_GRB200125B} and \ref{Duration_GRB200125B}. Peak fluxes of different time scales are naturally obtained in the same analysis, and the fluence is obtained by integrating the deconvolved flux history.

It is worth noting that, to avoid the failure of fit to each time bin, the parameters of GRB shape are fixed to the best model determined in joint fit to the time-averaged spectrum, while the amplitudes are set to be float. Therefore, in addition to the systematic effects mentioned in \cite{koshut1996}, an extra systematic error is introduced by fixed parameters. To estimate the uncertainty from the fixed parameters, we simulate tens of sets of parameters of the best GRB model by smearing the mean values with covariance matrix within one standard deviation. Then the same procedures are applied to calculate tens of sets of duration values. The maximum changes from the duration values with the mean value are taken to be the systematic error from fixed GRB parameters. The systematic uncertainties of peak fluxes and fluences from the same source are also estimated in the same way. Note that in most cases of spectral parameters that are well determined, the system uncertainties of this kind are less than or numerically comparable with the the systematic effects, which could be ignored.
As a sample of GRB 200125B, the total fluence (10-2000~keV) is 5.92e-05$\pm$4.54e-07 $\rm erg$ $\rm cm^{-2}$, the peak flux~($\rm ph$ $\rm cm^{-2}$ s$^{-1}$) in 10-2000~keV of timescales of 64~ms, 256~ms, 1024~ms is 175.35$\pm$9.57, 149.25$\pm$3.78, 110.27$\pm$1.63 respectively.

\section{CATALOG RESULTS} \label{sec:catalog}
Table~\ref{tab:main_table} lists the 322 searched bursts that are classified as GRBs. In the last column of Table~\ref{tab:main_table}, abbreviation of GRBs types are given. Among 322 searched bursts files, 202 `GOLDEN' samples are analyzed and collected in all. 44 `SILVER' , 50 `BRONZE' and 26 `IRON' samples are also listed in Table~\ref{tab:GRBsample}.

The results of the duration of GRBs are derived from joint analyses of HXMT with $\emph{Fermi}$/GBM, $\emph{Swift}$/BAT or GECAM discussed in \ref{subsec:DPF}, and shown in Table~\ref{tab:durations}. The values of $T_{\rm 50}$ and $ T_{\rm 90}$ in 10-2000~keV energy range are listed along with their respective statistical error
estimates and start times relative to the trigger time of HXMT. As part of the duration analysis, peak fluxes and fluence are computed. Table~\ref{tab:pf_fluence} shows the total fluence and peak fluxes of different time scales in 10-2000 keV. Table~\ref{tab:silbrodurations} lists the durations of `SILVER' and `GOLDEN' samples, which obtained from HXMT/CsI GRBs signal counts.

\section{DISCUSSION and SUMMARY} \label{sec:discussion}
\begin{figure*}
\begin{center}
\epsscale{.9}
\plotone{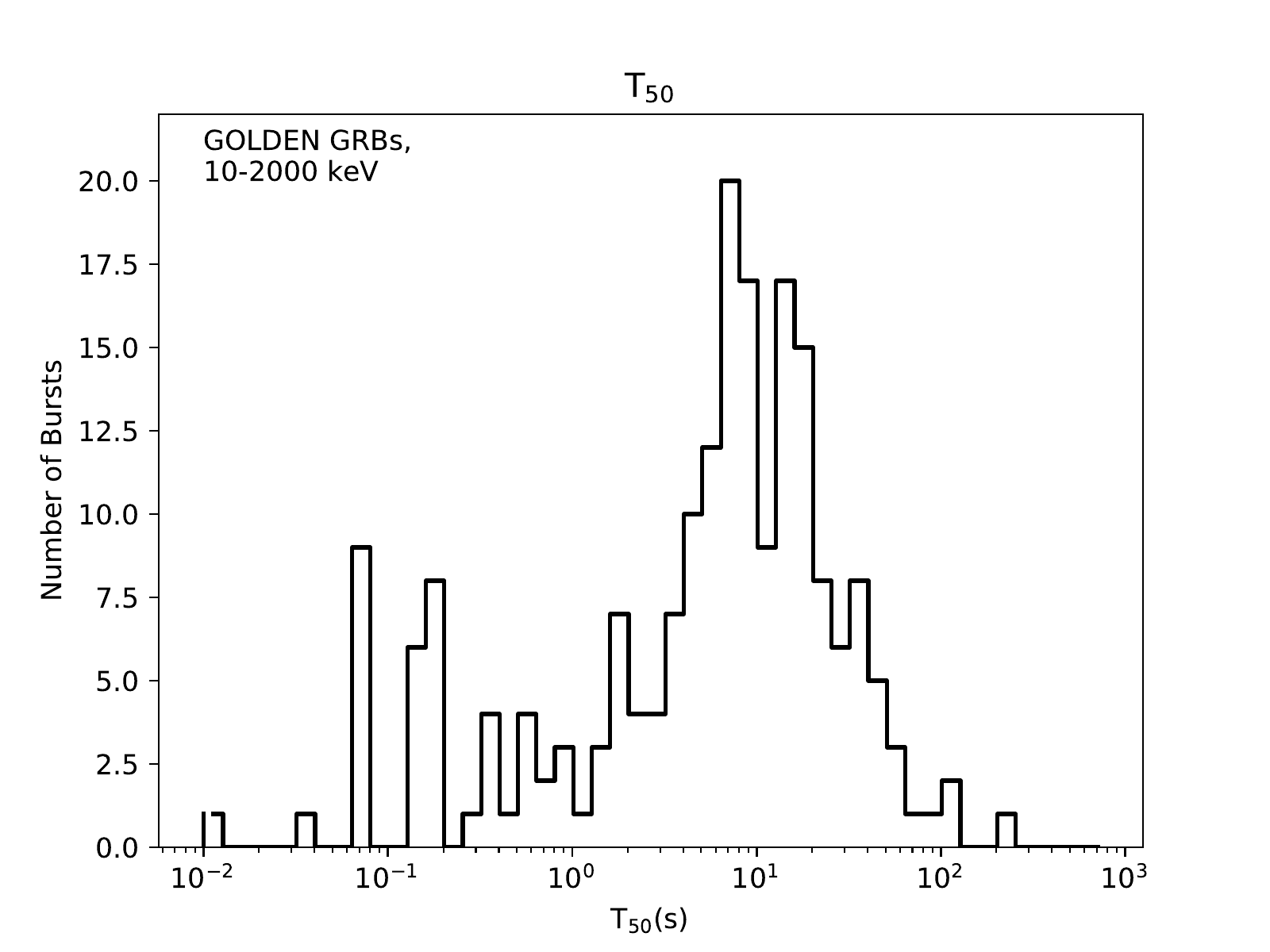}
\plotone{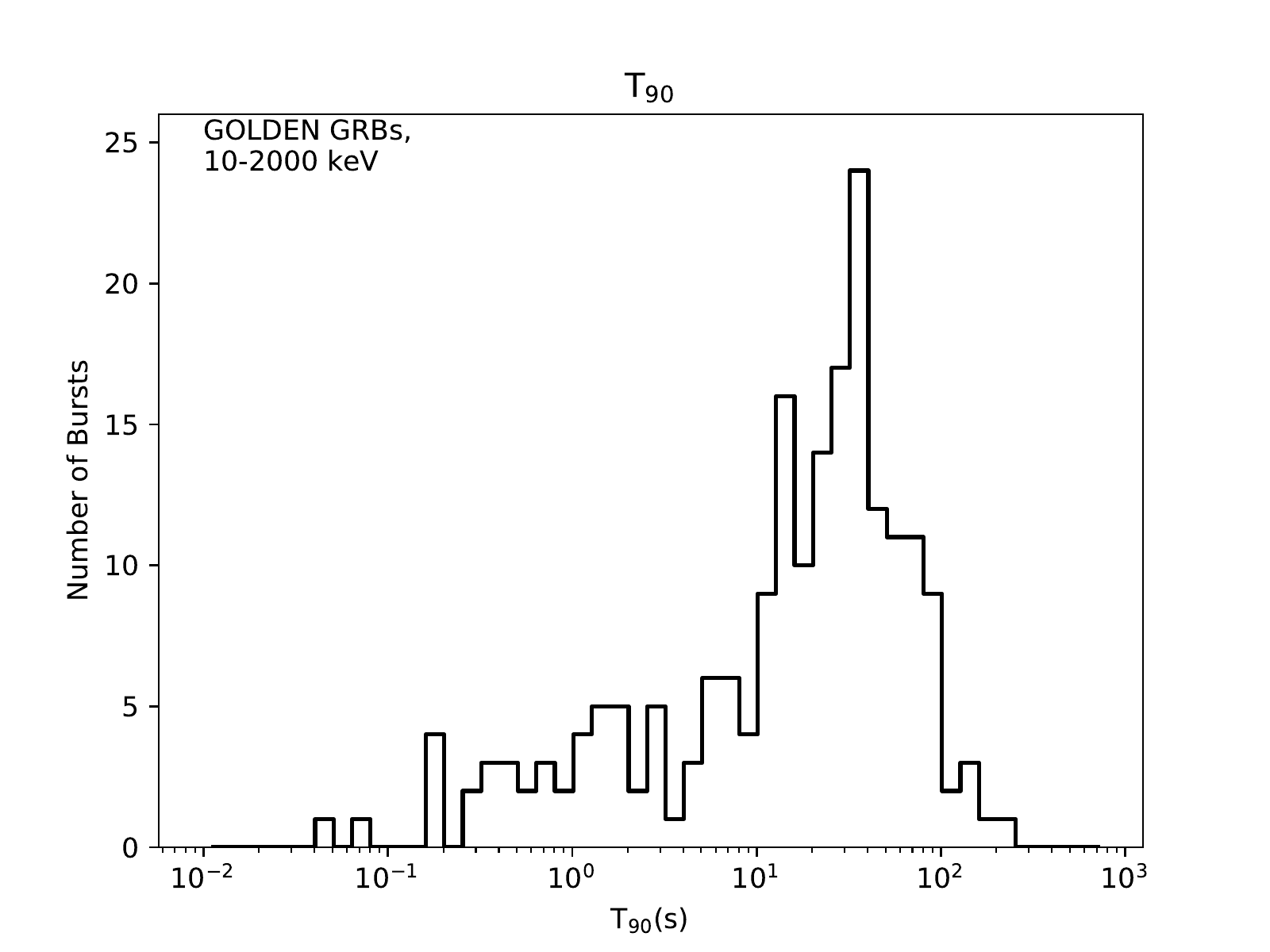}
\caption{\label{fig:Duration} Distributions of `GOLDEN' GRB durations in the 10-2000 keV energy range are shown. The upper plot shows $ T_{ \rm 50}$ and the lower plot shows $ T_{\rm 90}$.}
\end{center}
\end{figure*}

Histograms of the $ T_{\rm50}$ and $ T_{\rm90}$ distributions are shown in 10-2000~keV energy range in Figure~\ref{fig:Duration}. Using the conventional division between the short and long GRB classes ($ T_{\rm90}$=2~s), we find 35 (17\%) of the 202 `GOLDEN' GRBs to be in the short regime. Within the quoted duration errors, the number of short
GRB events ranges from 28 (14\%) to 39 (19\%). It is consistent with the results from \emph{Fermi}/GBM GRB catalogs covering ten years~\citep{von_Kienlin_2020}, which has 395 (17\%) short GRBs from statistics in 10-1000 keV. Due to the limited statistic of GRB numbers, we do not perform lognormal fits to the distribution of durations.

\begin{figure*}[!htbp]
\begin{center}
 \centering
  \includegraphics[width=0.7\textwidth]{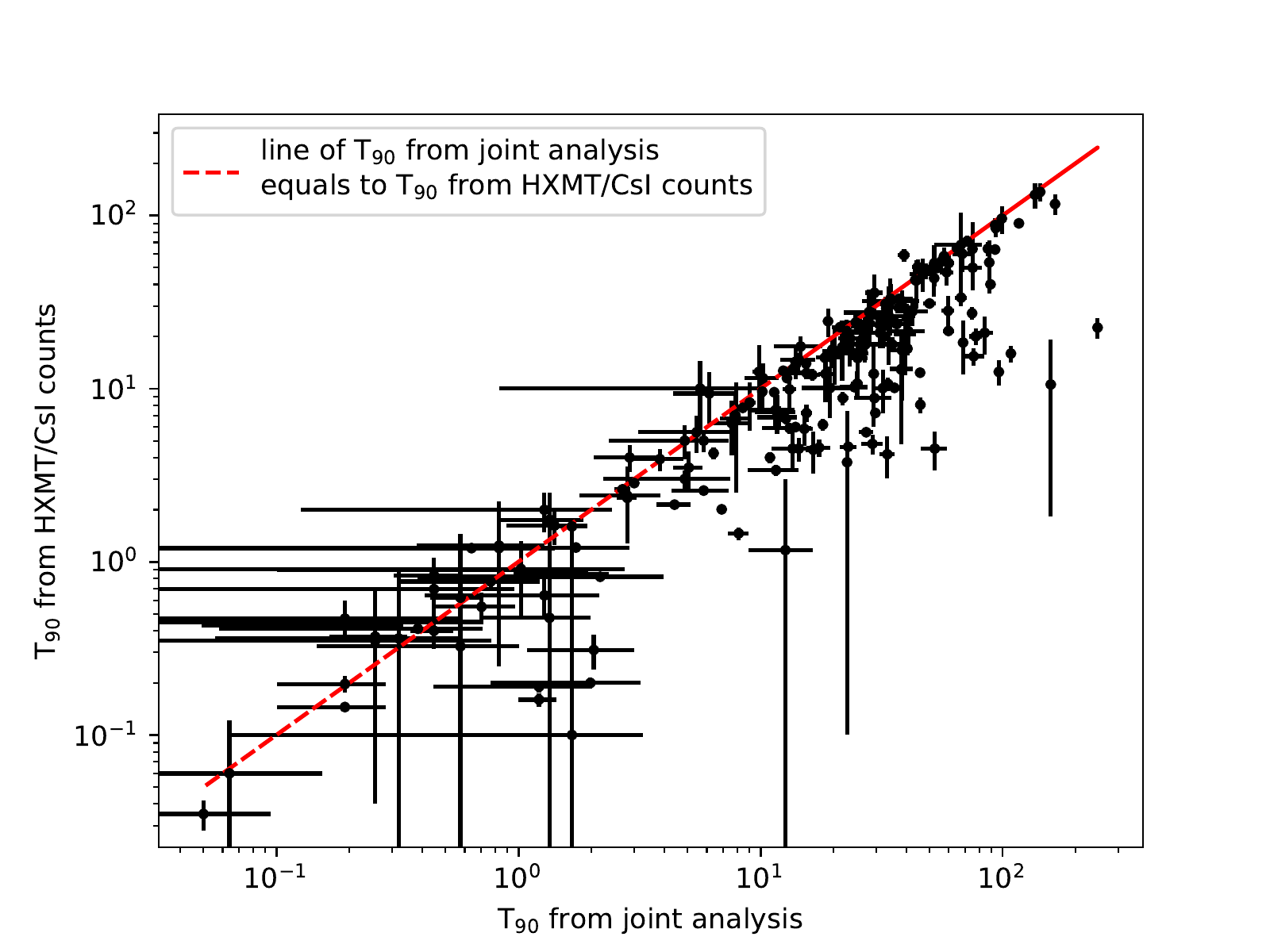}\\
\caption{\label{fig:Duration_compare} A comparison between the durations ($ T_{\rm 90}$) of `GOLDEN' GRBs from joint analysis and those obtained from HXMT/CsI GRBs counts are shown. The red line denotes the case of these two are equal to each other.}
\end{center}
\end{figure*}

Figure~\ref{fig:Duration_compare} shows the comparison of durations ($T_{\rm 90}$) from joint analyses with other missions and those from HXMT/CsI GRBs data, where the red line denotes the case of these two are equivalent to each other. \textbf{Most GRBs are below the red line, because energy range of HXMT/CsI is higher than those of other missions, such as \emph{Fermi}/GBM and that GRBs tend to be shorter at higher energy. Therefore, joint analysis is necessary for the measurement of duration in 10-2000 keV.} 

\begin{figure*}
\begin{center}
\epsscale{.9}
\plotone{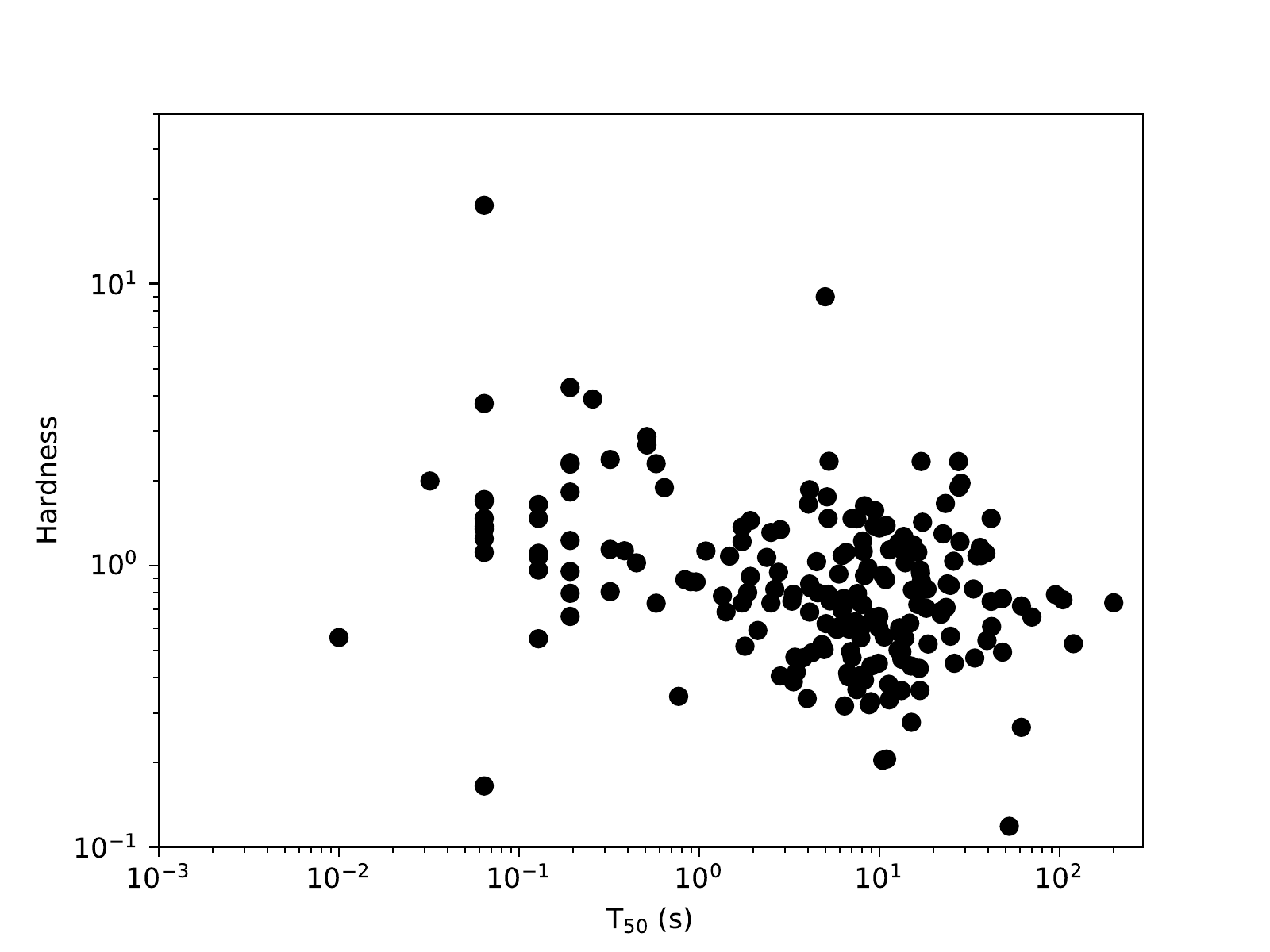}
\plotone{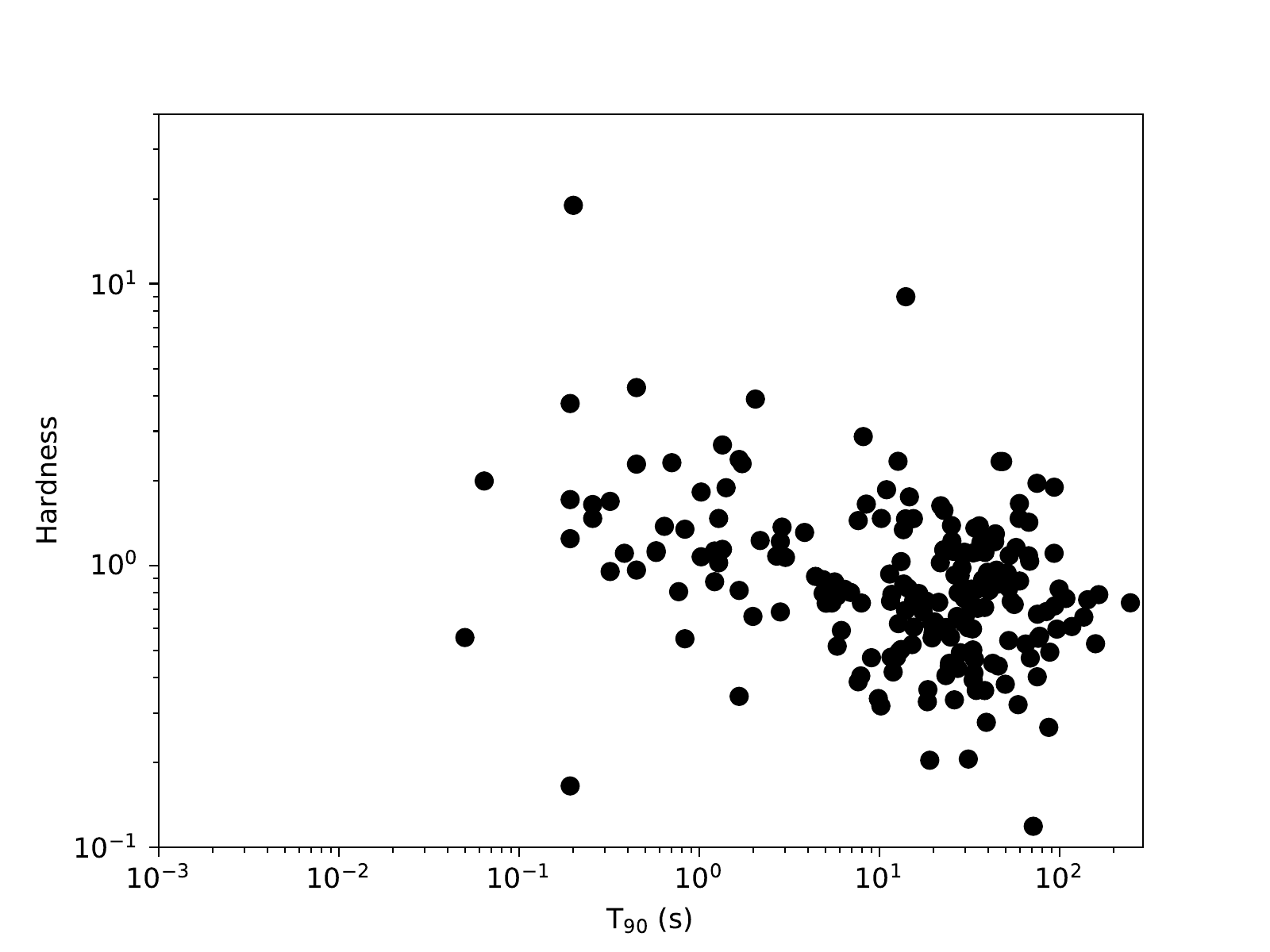}
\caption{\label{fig:Hardness_vs_Duration_12} Scatter plots of spectral hardness vs. duration are shown for the two duration measures $ T_{ \rm 50}$ (upper plot) and $ T_{\rm 90}$ (lower plot) as shown in Figure~\ref{fig:Duration}. The hardness is calculated as the ratio of the flux density from spectral parameters determined from time-averaged spectral fits in 50-300~keV to that in 10-50~keV. For clarity, the estimated errors are not shown but can be quite large for the weak events. Nevertheless, the anti-correlation of spectral hardness with burst duration is evident. }
\end{center}
\end{figure*}

 Despite of the limited statistic of `GOLDEN' GRBs, as shown in Figure~\ref{fig:Hardness_vs_Duration_12}, anti-correlation of spectral hardness with duration is confirmed by the relationship between the hardness and duration, where the hardness is calculated as the ratio of the flux density from spectral parameters determined from time-averaged spectral fits in 50-300~keV to that in 10-50~keV. We note that, here the durations are those from joint analyses in energy range of 10-2000~keV, which also illustrates the anti-correlation between the hardness of GRBs spectra and duration.

\begin{figure*}
\begin{center}
\epsscale{.8}
\plotone{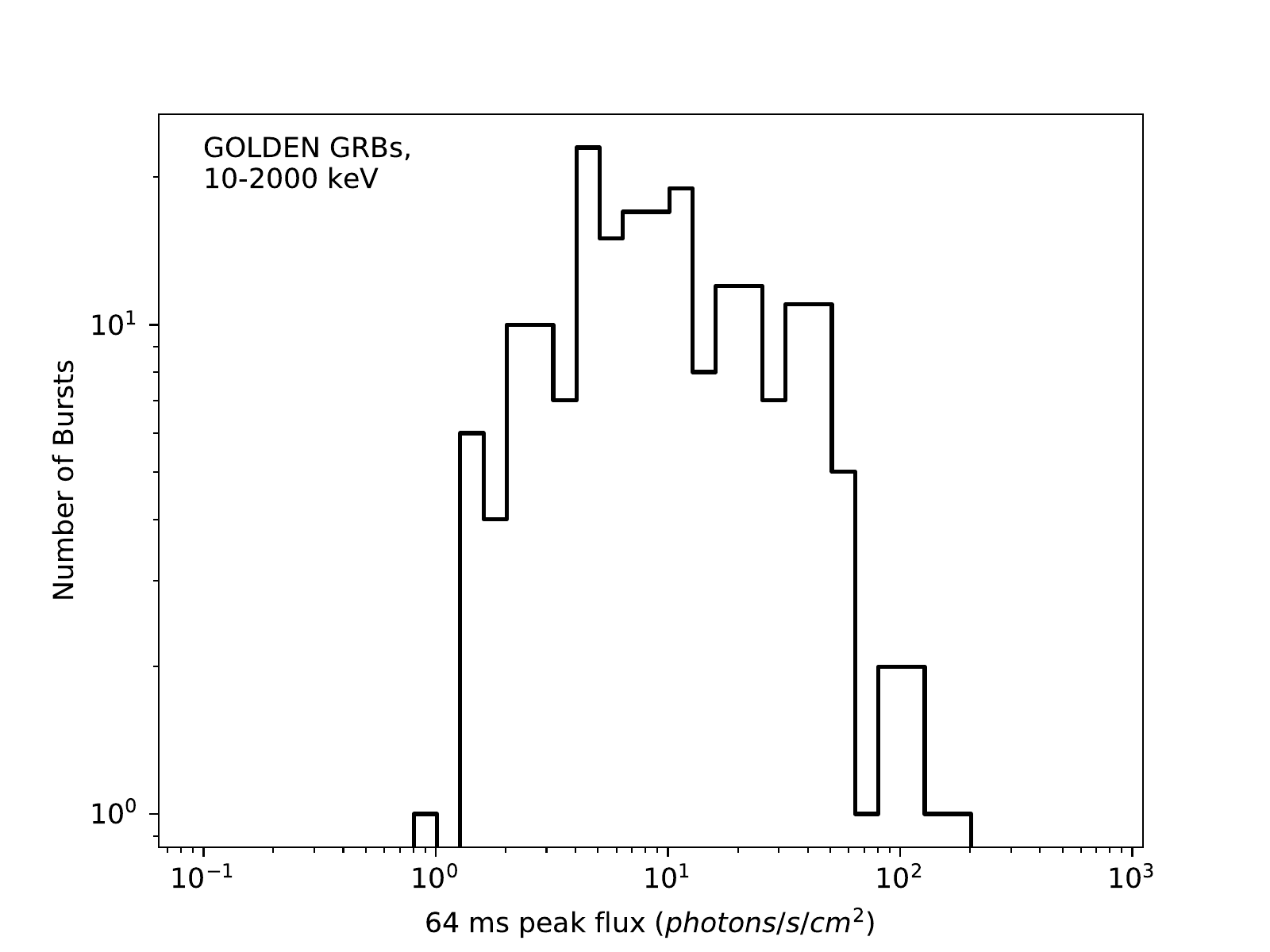}
\caption{\label{fig:Distribution peakflux_64ms} Distribution of `GOLDEN' GRBs peak flux on the 0.064 s timescale is shown in energy range of 10-2000~keV.}
\end{center}
\end{figure*}

\begin{figure*}
\begin{center}
\epsscale{.8}
\plotone{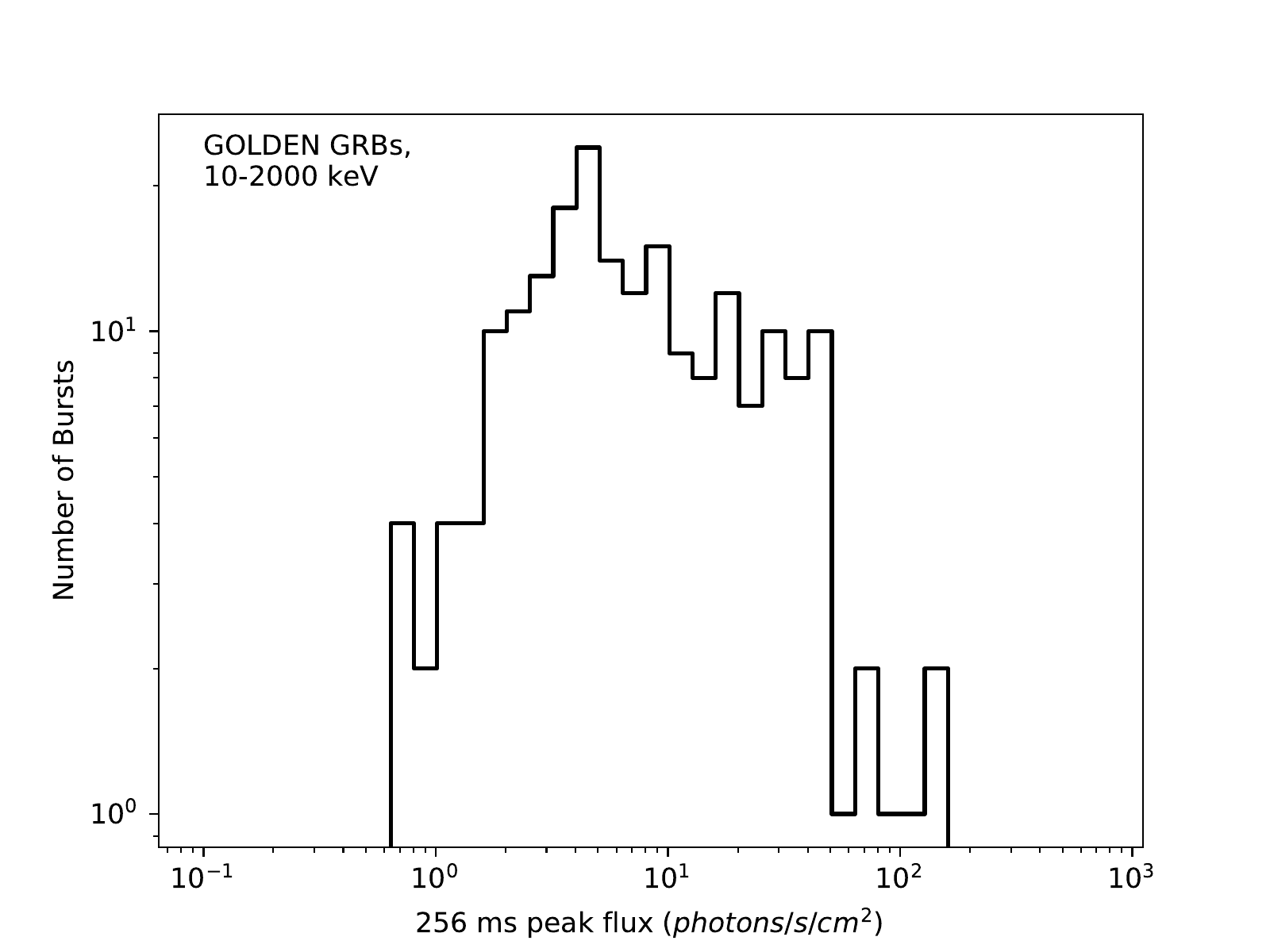}
\caption{\label{fig:Distribution peakflux_256ms} Distribution of `GOLDEN' GRBs peak flux on the 0.256 s timescale is shown in energy range of 10-2000~keV.}
\end{center}
\end{figure*}

\begin{figure*}
\begin{center}
\epsscale{.8}
\plotone{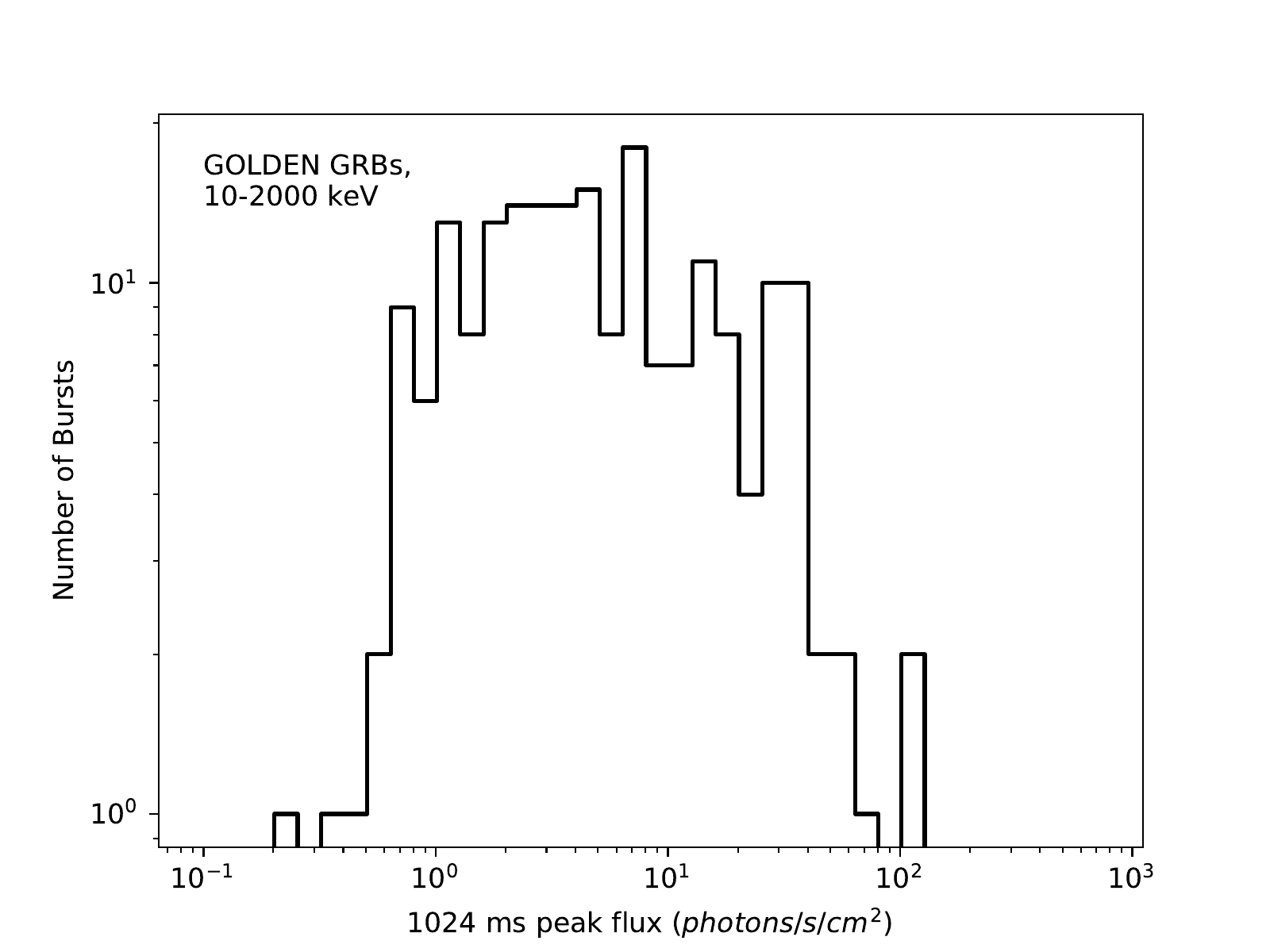}
\caption{\label{fig:Distribution peakflux_1024ms} Distribution of `GOLDEN' GRBs peak flux on 1.024 s timescale is shown in energy range of 10-2000~keV.}
\end{center}
\end{figure*}

\begin{figure*}
\begin{center}
\epsscale{.9}
\plotone{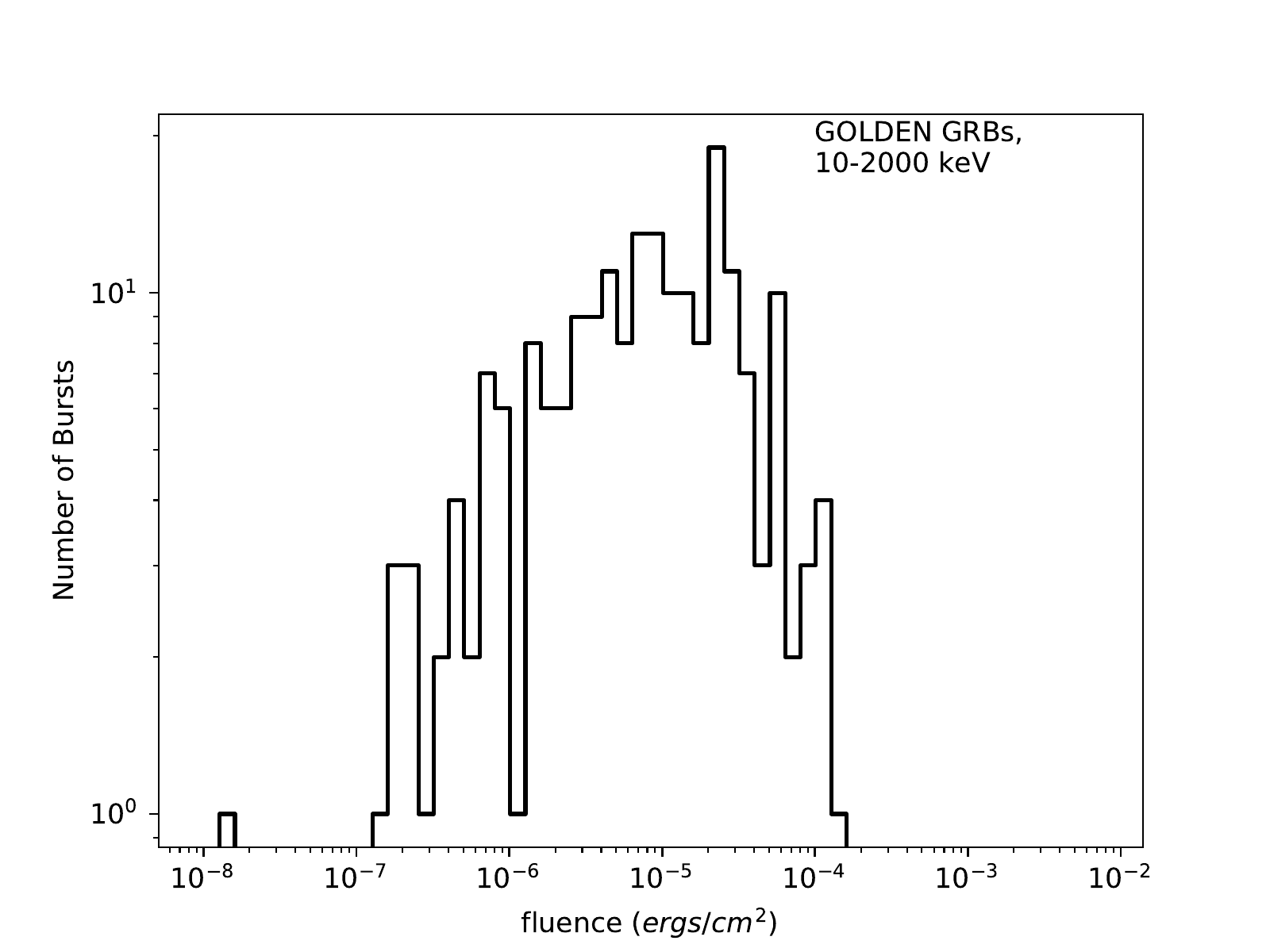}
\caption{\label{fig:Distribution fluence} Distribution of `GOLDEN' GRBs fluence is shown in energy range of 10-2000~keV.}
\end{center}
\end{figure*}

 Distributions of GRBs fluence are shown in Figure~\ref{fig:Distribution fluence}. Peak fluxes on timescales of 64~ms, 256~ms, 1024~ms are shown in Figure~\ref{fig:Distribution peakflux_64ms}, Figure~\ref{fig:Distribution peakflux_256ms} and Figure~\ref{fig:Distribution peakflux_1024ms} respectively.  These distributions are not consistent with those from \cite{von_Kienlin_2020}. The contributions are very small both below 10$^{-7}$~$\rm erg$ $\rm cm^{-2}$ and 1~$\rm ph$ $\rm cm^{-2}$ s$^{-1}$ in distributions of fluence and peak fluxes. It could be explained by the sharply fall of effective area below 100~keV. Weak short GRBs cannot cause enough statistics in HXMT/CsI, and thus cannot be detected significantly. Extremely bright GRBs would not be included in `GOLDEN' samples because they are very likely to cause data saturation in HXMT and be labeled as `IRON' samples.



\acknowledgements {
This work made use of the data from the \emph{Insight-}HXMT mission, a project funded by China National Space Administration (CNSA)
and the Chinese Academy of Sciences (CAS). The authors thank supports from the National Program on Key Research and Development Project
(Grants No. 2016YFA0400801, 2021YFA0718500), the National Natural Science Foundation of China under Grants No. U1838113,
and the Strategic Priority Research Program on Space Science, the Chinese Academy of Sciences (Grant No. XDB23040400, XDA15052700). The authors are very grateful to the public data of \emph{Fermi}/GBM, \emph{Swift}/BAT and GECAM.
}
\clearpage



 commands
\end{deluxetable*}

\appendix

\section{light curves and spectra of GRB 200125B and GRB 210121A }\label{sec:sample}
\label{sec:sample}
\begin{figure*}[!htbp]
  \centering
  \includegraphics[width=0.8\textwidth]{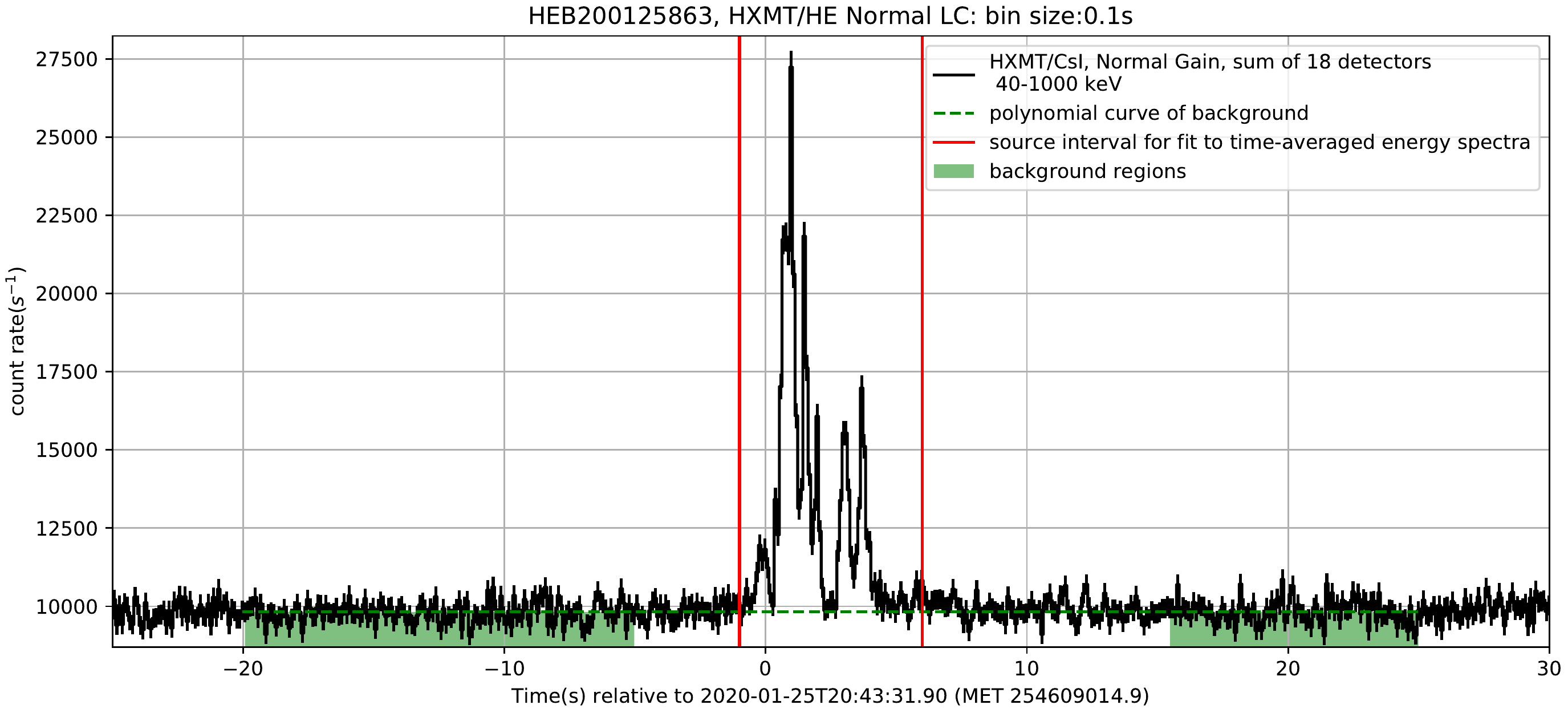}\\
  \caption{Light curve of GRB 200125B (HEB200125863) of 18 CsI detectors. Vertical lines
indicate the regions selected for the fit to the time-averaged spectra. Green regions defines the background region. Here MET means the mission elapsed time of HXMT.
  }\label{LC_GRB200125B}
\end{figure*}

\begin{figure*}[!htbp]
  \centering
  \includegraphics[width=0.7\textwidth]{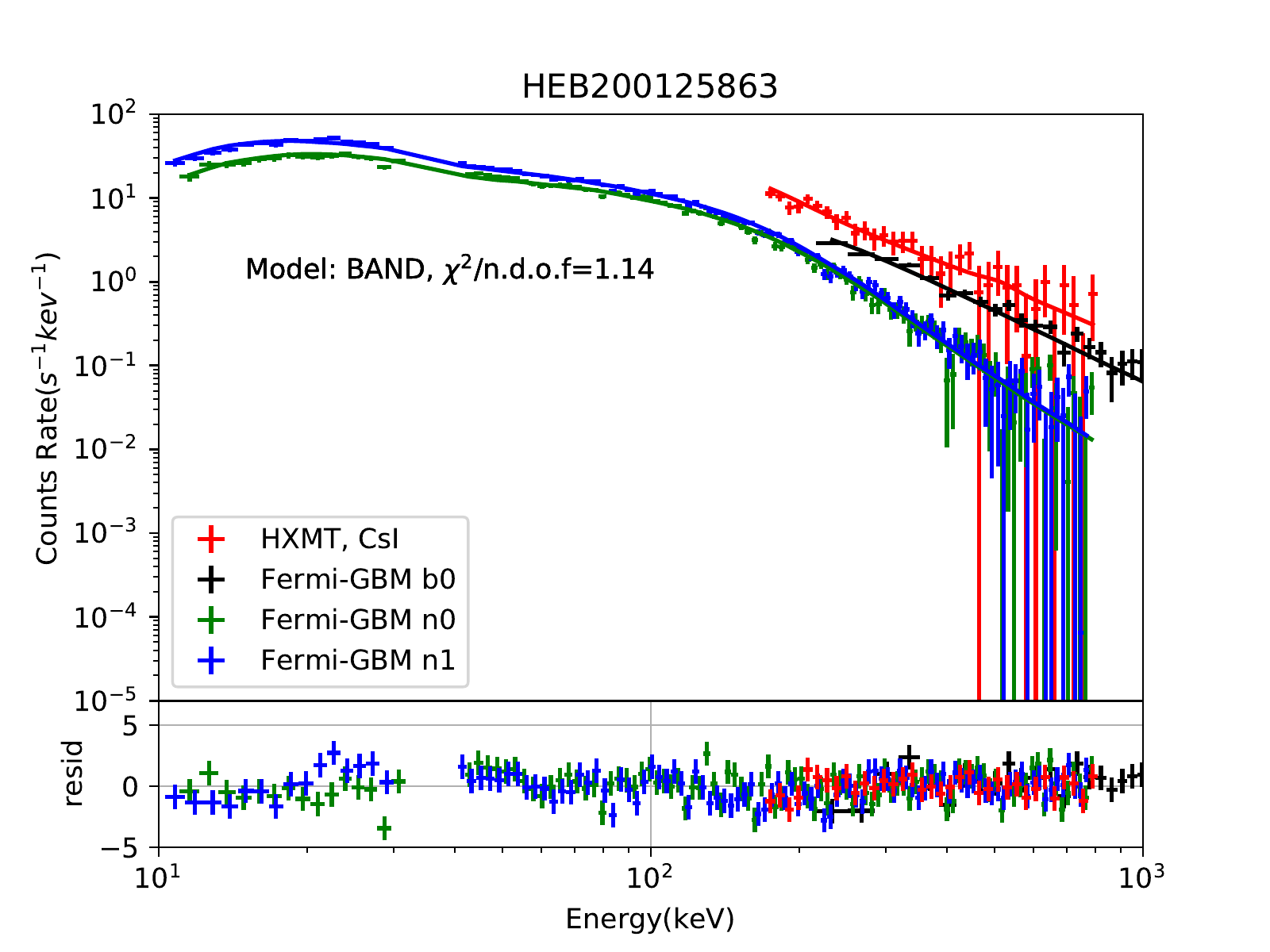}\\
  \caption{ Joint analysis of time-averaged spectrum of GRB 200125B (HEB200125863). Data from 18 HXMT/CsI detectors and \emph{Fermi}/GBM NaI detector 0, 1, BGO detector 0 are utilized.
  }\label{spectrum_GRB200125B}
\end{figure*}

\begin{figure*}[!htbp]
  \centering
  \includegraphics[width=0.7\textwidth]{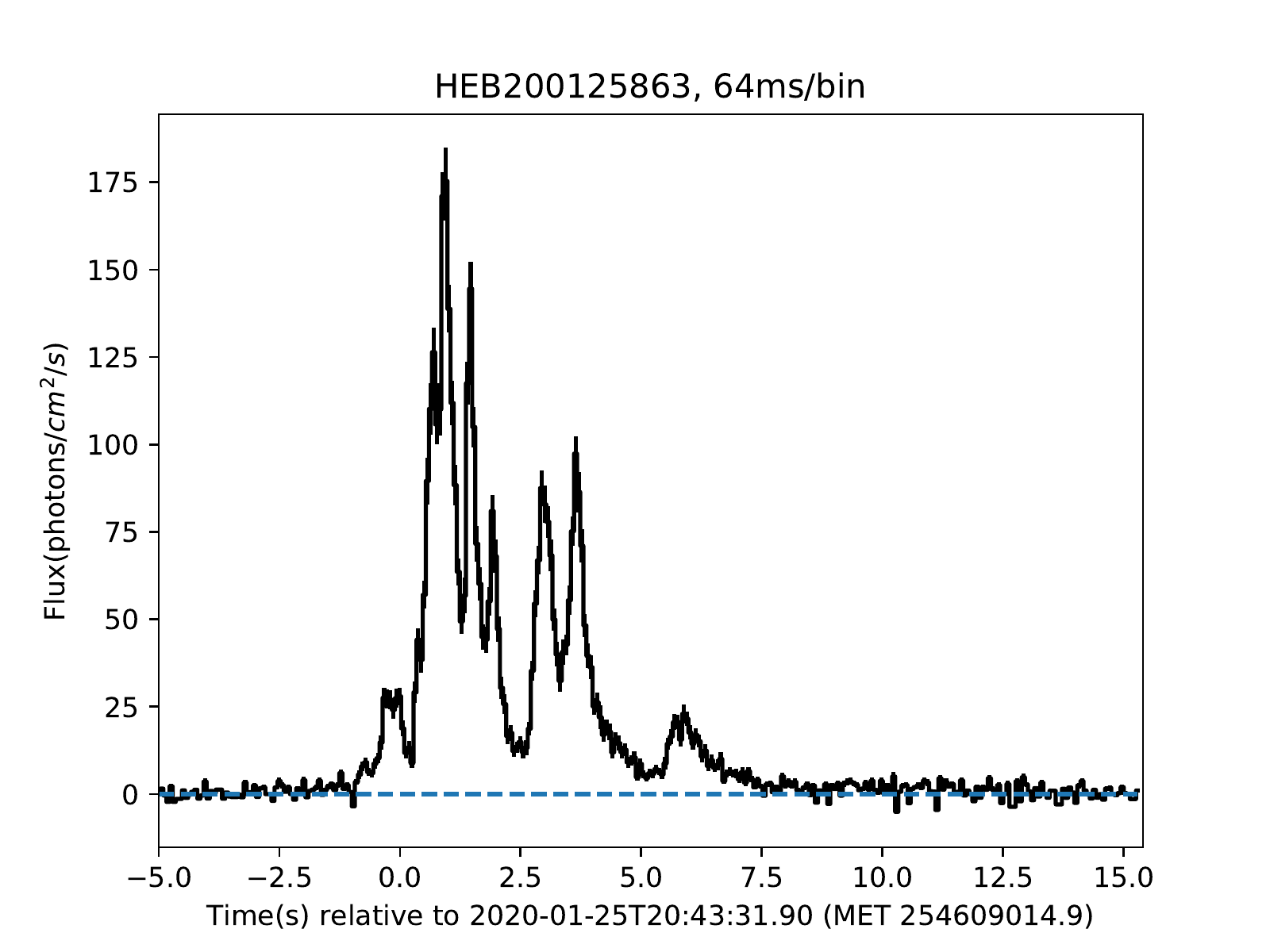}\\
  \caption{Photon flux light curve for GRB 200125B (HEB200125863) produced by the duration analysis.
Data from 18 HE/CsI detectors and \emph{Fermi}/GBM NaI detector 0, 1, BGO detector 0 are used. Here MET means the mission elapsed time of HXMT.
  }\label{flux_GRB200125B}
\end{figure*}

\begin{figure*}[!htbp]
  \centering
  \includegraphics[width=0.7\textwidth]{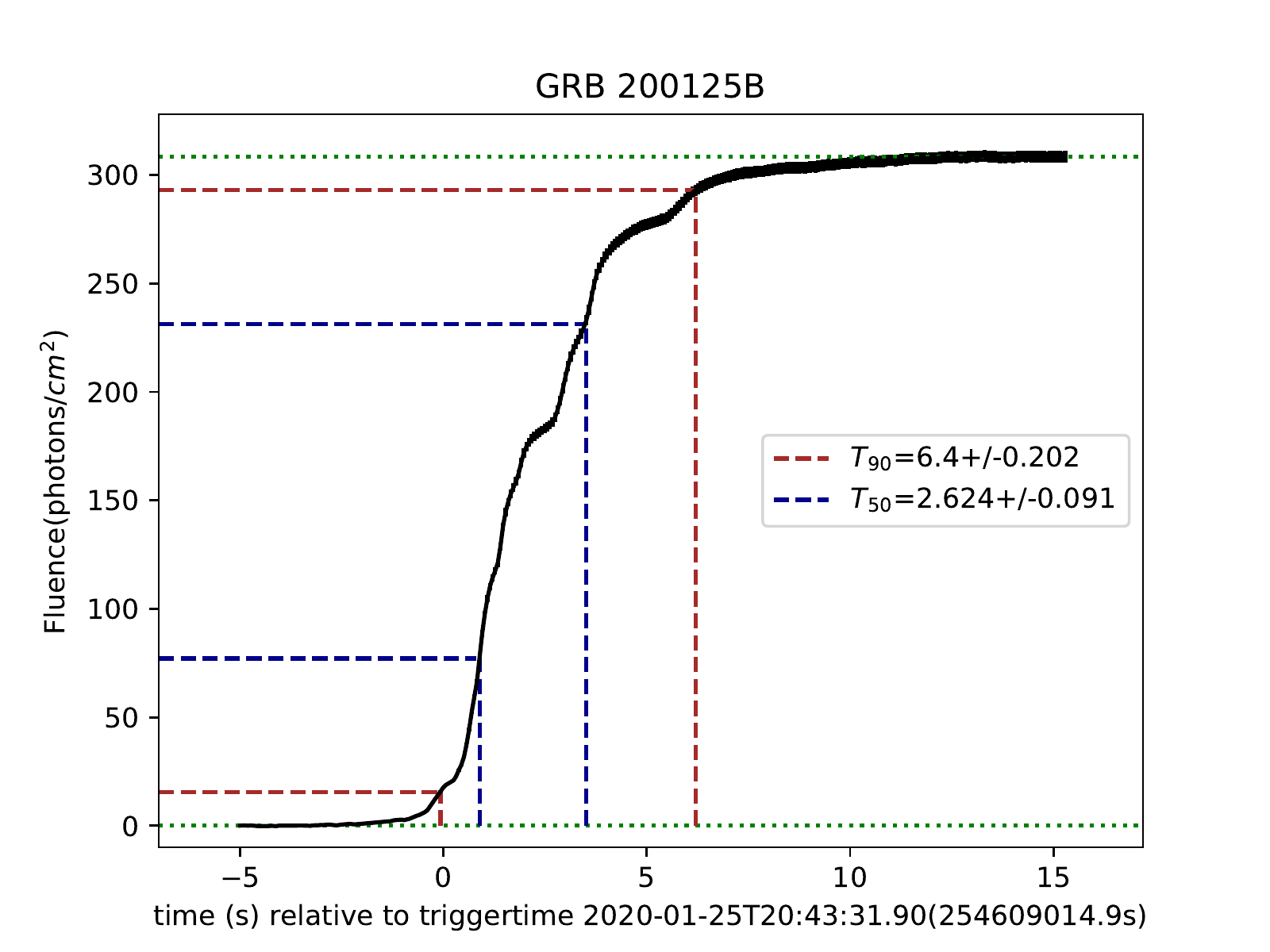}\\
  \caption{The duration plot for GRB 200125B (HEB200125863). Horizontal dashed lines in dark red are drawn
at $5\%$ and $95\%$ of the total fluence, while those in dark blue at $25\%$ and $75\%$. Vertical dotted lines are drawn at the times
corresponding to those same fluences, thereby defining the $ T_{ \rm 50}$ and $ T_{ \rm 90}$ intervals. The green dotted horizontal lines are corresponding to the $0\%$ and $100\%$ of the total integration of the flux respectively. Here MET means the mission elapsed time of HXMT.
  }\label{Duration_GRB200125B}
\end{figure*}

\begin{figure*}[!htbp]
  \centering
  \includegraphics[width=0.7\textwidth]{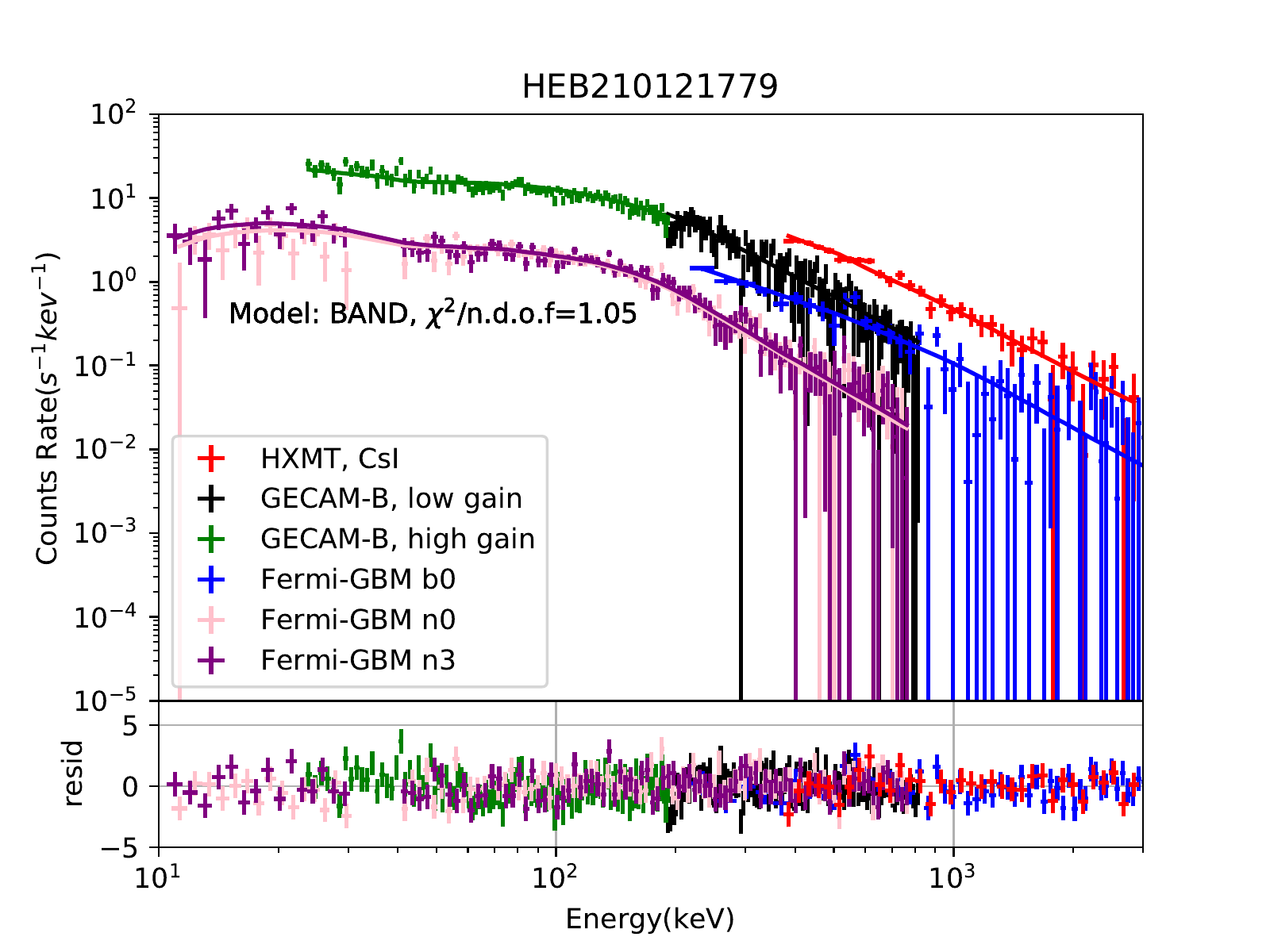}\\
  \caption{ Joint analysis of time-averaged spectrum of GRB 210121A (\textbf{HEB210121779}). Data from 18 HE/CsI detectors and \emph{Fermi}/GBM NaI detector 0, 3, BGO detector 0 are utilized.
  }\label{spectrum_GRB210121A}
\end{figure*}

\begin{figure*}[!htbp]
  \centering
  \includegraphics[width=0.7\textwidth]{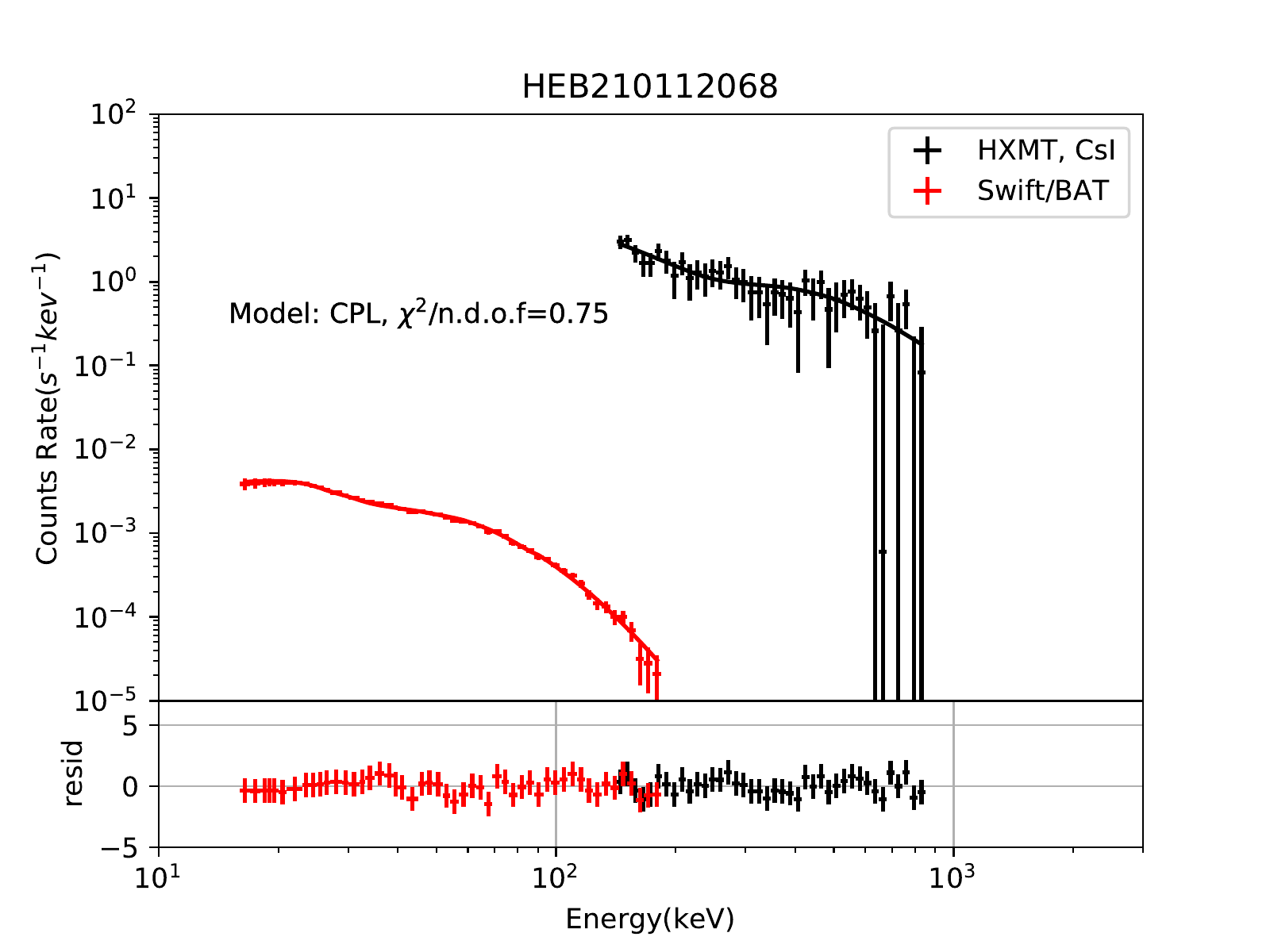}\\
  \caption{ Joint analysis of time-averaged spectrum of GRB 210112A (HEB210112068). Data from 18 HE/CsI detectors and \emph{Swift}/BAT are utilized.
  }\label{spectrum_GRB210112A}
\end{figure*}
\clearpage
\bibliographystyle{aasjournal}
\bibliography{catalogbib}
\end{document}